\documentclass[letterpaper,twocolumn,10pt]{article}

\newif\ifSPACEHACK
\SPACEHACKfalse

\newif\ifDEBUG
\DEBUGtrue

\newif\ifANONYMOUS
\ANONYMOUSfalse

\newif\ifTECHREPORT
\TECHREPORTtrue

\usepackage{usenix}

\usepackage{cleveref}
\crefformat{section}{\S#2#1#3}
\crefname{figure}{Figure}{Figures}
\crefname{table}{Table}{Tables}
\crefname{listing}{Listing}{Listings}
\crefname{theorem}{Theorem}{Theorems}
\crefname{thm}{Theorem}{Theorems}
\crefname{lemma}{Lemma}{Lemmata}
\crefname{equation}{Eqt.}{Eqts.}
\crefformat{Grammar}{Grammar #1}
\crefname{paragraph}{paragraph}{paragraphs}
\crefname{appendix}{Appendix}{Appendices}
\Crefname{paragraph}{Paragraph}{Paragraphs}
\newcommand{\inlinequote}[1]{``\textit{#1}''}

\newenvironment{myquote}%
  {\list{}{\leftmargin=0.12in\rightmargin=0.1in}\item[]}%
  {\endlist}
\newcommand{\blockquote}[1]{\begin{myquote}``\textit{#1}''\end{myquote}}

\newcommand{\myparagraph}[1]{\paragraph{\textbf{#1}}}
\renewcommand{\myparagraph}[1]{\textbf{#1}}

\newcommand{\mysubparagraph}[1]{\indent\emph{\underline{#1}}}

\renewcommand{\arraystretch}{1.2}

\usepackage{subcaption}
\usepackage{graphicx}
\usepackage{tcolorbox}
\usepackage{tabularx}
\usepackage{float}
\usepackage{url}

\usepackage{breakurl}
\usepackage{xr}
\usepackage{xspace}
\usepackage{booktabs}
\usepackage{hyperref}

\newcommand{\ie}{\textit{i.e.,\ }}
\newcommand{\eg}{\textit{e.g.,\ }}
\newcommand{\etal}{\textit{et al.\ }}
\newcommand{\etals}{\textit{et al.}'s\ }
\newcommand{\adhoc}{\textit{ad hoc\xspace}}
\newcommand{\tahiti}{TaHiTI\xspace}
\newcommand{\attack}{ATT\&CK\xspace}

\ifDEBUG
\newcommand{\tm}[1]{\textcolor{red}{Trey: #1}}
\newcommand{\JD}[1]{\textcolor{blue}{Jamie: #1}}
\newcommand{\PJ}[1]{\textcolor{orange}{Purvish: #1}}
\else
\newcommand{\tm}[1]{}
\newcommand{\JD}[1]{}
\newcommand{\PJ}[1]{}
\fi

\newcommand{\new}[1]{\textcolor{teal}{#1}}
\renewcommand{\new}[1]{#1}

\usepackage{dblfloatfix}

\usepackage{multirow}
\usepackage{graphicx}

\usepackage{ulem}
\newcommand{\NOTABLEREMOVAL}[1]{\sout{#1}}
\usepackage{enumitem}

\newcommand{\maxquote}{5\xspace}

\begin{document}


\title{\Large \bf An Interview Study on Third-Party Cyber Threat Hunting Processes \\ in the U.S. Department of Homeland Security}

\ifANONYMOUS
\author{
{Anonymous author(s).}
}
\else


\author{
{\rm William P. Maxam III}\\
United States Coast Guard Academy\thanks{Work performed while at Purdue University.}
\and
{\rm James C. Davis}\\
Purdue University
} 

\fi

\maketitle

\begin{abstract}
Cybersecurity is a major challenge for large organizations.
Traditional cybersecurity defense is reactive. 
Cybersecurity operations centers keep out adversaries and incident response teams clean up after break-ins.
Recently a proactive stage has been introduced: Cyber Threat Hunting (TH) looks for potential compromises missed by other cyber defenses.
TH is mandated for federal executive agencies
and government contractors.
As threat hunting is a new cybersecurity discipline, most TH teams operate without a defined process.
The practices and challenges of TH have not yet been documented. 

To address this gap, this paper describes the first interview study of threat hunt practitioners.
We obtained access and interviewed 11 threat hunters associated with the U.S. government's Department of Homeland Security.
Hour-long interviews were conducted.
We analyzed the transcripts with process and thematic coding. 
We describe the diversity among their processes, show that their processes differ from the TH processes reported in the literature, and unify our subjects' descriptions into a single TH process.
We enumerate common TH challenges and solutions according to the subjects.
The two most common challenges were difficulty in assessing a Threat Hunter's expertise, and developing and maintaining automation. 
We conclude with recommendations for TH teams (improve planning, focus on automation, and apprentice new members) and highlight directions for future work (finding a TH process that balances flexibility and formalism, and identifying assessments for TH team performance).

\end{abstract}



\section{Introduction}

Computer network security is a challenge in the modern world.
Cyber intrusions are a concern for both governments and private corporations.
Unauthorized network infiltrations cost individual organizations an average of \$13 million a year~\cite{noauthor_cost_2019} and may compromise their operations or intellectual property. 
Governments have additional non-monetary concerns, such as protecting election systems and maintaining national security~\cite{noauthor_national_2018}. 
The longer an adversary dwells undetected on a network, the more damage the adversary can cause.
One analysis found that a 50\% reduction in dwell time would reduce the cost of an attack by $\sim$30\%~\cite{brink_quantifying_2017}. 
IBM found that data breaches cost on average \$1.12 million more if not contained within the first 200 days, 
with costs including lost revenue, regulatory and legal fees, and forensics activities~\cite{noauthor_cost_2020}.
They estimate the average adversary-dwell-time at 230 days, not including additional time to respond to the breach~\cite{noauthor_cost_2020}.

The primary method of finding undetected network intruders is a Cyber Threat Hunt (TH).
Threat Hunting is \inlinequote{a focused and iterative approach to searching out, identifying, and understanding adversaries internal to the defender's networks}~\cite{lee_who_2016}.
The government routinely performs \textbf{third-party hunting}, \ie TH on contractors' networks~\cite{noauthor_national_2018}.
The private sector conducts threat hunting as well~\cite{noauthor_national_2023}, more frequently internal hunting within their own organization.
According to a 2017 SANS Institute survey, across sectors including telecommunications, technology, government, healthcare, and finance, most organizations engage in threat hunting but in an immature way~\cite{lee_hunter_2017}.
Less than half of the organizations had a defined TH process~\cite{lee_hunter_2017}. 

The processes currently used by TH teams are not well documented~\cite{van_os_tahiti_2018}.
Prior research on TH teams does not describe TH processes in detail~\cite{wafula_carve_2019,agarwal_cyber_2021,bynum_cyber_2019,trent_modelling_2019} or is focused on internal hunt teams~\cite{van_os_tahiti_2018}.
This knowledge gap limits a team's ability to adopt best practices and improve their processes over time, and it also limits how well researchers can assist TH teams.

To address the lack of TH process understanding, we interviewed professional threat hunters.
We used a semi-structured interview methodology,
selected as a result of the data available from previous work done in the TH domain and an anticipated small sample size (\cref{design}). 
We interviewed 11 TH practitioners from two organizations within the US Department of Homeland Security (DHS).
Each subject was interviewed for $\sim$1 hour.
We combined subjects' TH process sketches and dialogue to create a unified process model of the US DHS threat hunt process.
We also thematically coded the transcripts to elicit subjects' problems and solutions.

Our results provide researchers with a better understanding of TH processes. 
We provide the first academic description of the cybersecurity landscape to include TH teams (\cref{background}) and the first published process model describing DHS TH teams (\cref{fig:CoarseInducedDiagram}).
Unlike prior literature that recommends a hypothesis-driven process, we report that the studied TH teams incorporate a \textit{data-driven} process.
We also document the challenges that practitioners report with this process, and describe the best practices they recommend (\cref{rq2_table}).
Their open questions are captured as future work (\cref{dis}).

Our contributions are:
\begin{itemize}[itemsep=0pt]
    \item 
    We provide an updated description of the cybersecurity landscape that includes Threat Hunting teams (\cref{background}).
    \item We characterize US DHS Threat Hunt processes, both in comparison to prior work and via a novel process induction
    \ifTECHREPORT
    (\cref{rq1}).
    \cref{fig:CoarseInducedDiagram} summarizes,~\cref{fig:DetailedInducedDiagram} details. 
    \fi
    \item We describe problems, possible solutions, and open questions discussed by TH practitioners (\cref{rq2,rq2_table}). 
\end{itemize}


\section{Background and Related Work}
\label{background}

In~\cref{layers} we describe the landscape of current cybersecurity defenses and the role of Threat Hunting (TH).
\cref{orgs} summarizes what is known about public and private TH teams. 
\cref{back_frame} describes common TH frameworks and processes. 
As there is little academic literature on Threat Hunt, this section is expository.
We rely in part on reputable ``grey literature''.

\subsection{Layers of cybersecurity defense}
\label{layers}
\cref{overview} illustrates typical layers of cybersecurity defense as they interact with an adversary.
We discuss each layer in turn. 

\ifTECHREPORT
    \begin{figure*}[ht]
    \includegraphics[width=1.75\columnwidth]{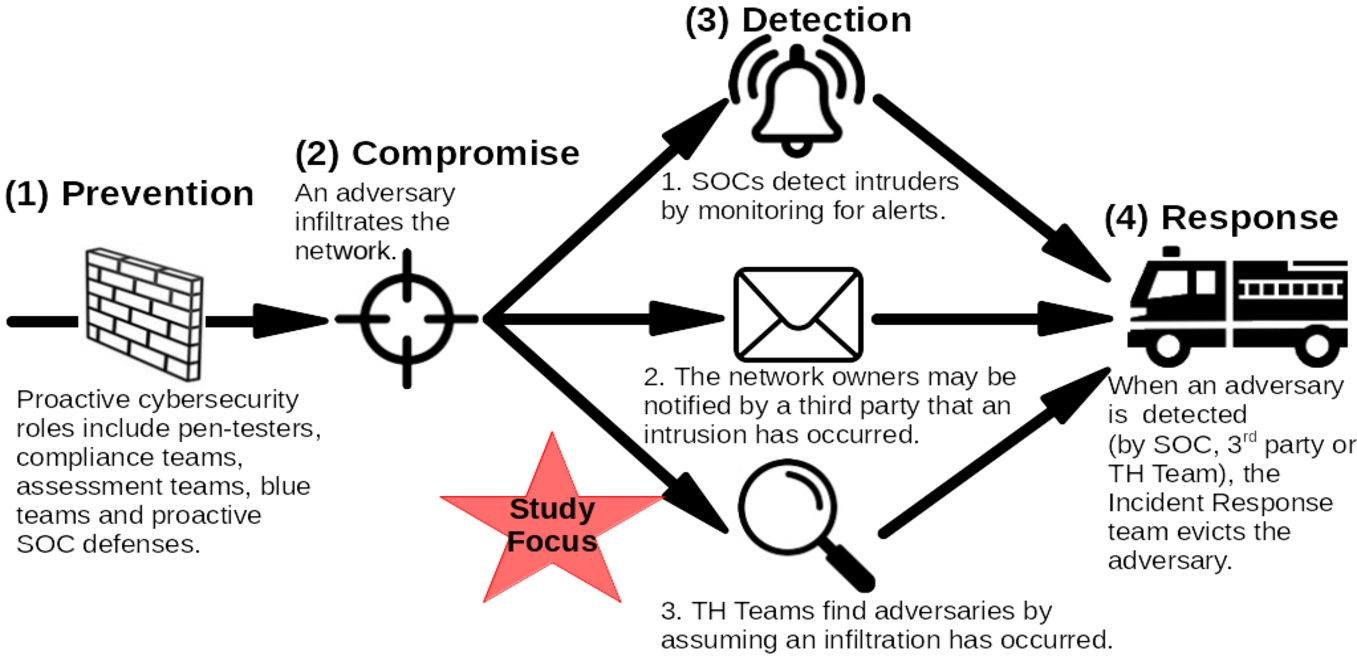}
    \centering
    \caption{
      Threat hunt is one of three common ways to discover an adversary once they have circumvented cyber defenses.
      Once an adversary is discovered, an IR team responds. 
      }
    \label{overview}
    \end{figure*}
\fi

\subsubsection{Prevention (Diagram Step 1)}
\label{prevention}
As~\cref{overview} shows, \textit{preventative} cybersecurity teams are the first line of defense against adversaries. 
These teams include traditional defensive teams such as Blue Teams, Compliance Teams, and Security Operations Centers (SOCs).
Blue teams seek to protect the network from intruders by improving the network's security~\cite{white_btfm_2017}.
Compliance Teams enforce cybersecurity best practices across a network~\cite{donalds_cybersecurity_2020}.
SOCs are primarily responsible for detecting adversaries as they attempt to gain access to a network~\cite{alharbi_security_nodate}.
SOCs in particular have received much attention from researchers.
Researchers have interviewed SOC members~\cite{cho_capturing_2020} and embedded in SOCs~\cite{sundaramurthy_tale_2014}.
Classification systems exist for SOCs to measure their maturity and capability~\cite{jacobs_classification_2013}, best practices have been enumerated~\cite{majid_success_2019} and SOC processes are being studied~\cite{sacher-boldewin_intelligent_2022}.

There are also offensive preventative security teams.
They include Red Teams and Penetration Testing Teams~\cite{everson_network_2020}.
These teams take offensive actions in order to simulate adversaries.
Penetration Testing Teams typically look only for vulnerabilities at the network edge, while red teams seek to infiltrate past the network edge and further into the network.

\subsubsection{Compromises (Diagram Step 2)}
\label{compromise}
In Step 2 of~\cref{overview}, the adversary evades these countermeasures and \textit{compromises} the network. 
If an adversary compromises a network without detection, they can cause significant damage~\cite{brink_quantifying_2017}.
The longer adversaries remain undetected, the more costly the intrusion becomes~\cite{brink_quantifying_2017,noauthor_cost_2020}.
Attackers constantly innovate their capabilities, and many works have described compromises such as SolarWinds\cite{alkhadra_solar_2021}, Stuxnet\cite{matrosov_stuxnet_2010}, and the 2015 Ukrainian power grid attack\cite{whitehead_ukraine_2017}.

\subsubsection{Detection (Diagram Step 3)}
\label{detection}

Step 3 of~\cref{overview} shows three ways in which such an adversary may subsequently be \textit{detected}.
The upper route depicts the SOC detecting the adversary, \eg via one of the SOC's internal network or host-based sensors.
The middle route shows a notification by an agency such as the US FBI.
The third path is \textit{proactive discovery}, \ie Threat Hunt, our study's focus.

TH teams perform a unique function, hunting for adversaries internal to the network boundary. 
To identify undetected adversaries on the network, TH teams search for adversaries that have evaded detection by usual methods~\cite{lee_who_2016}.
This is analogous to a military unit relying both on gate guards (the SOC) and also patrols inside the camp to ``hunt'' for adversaries that breach the gate. 
TH teams do not configure network defences (as blue teams and compliance teams do), nor do they take offensive action (as red teams and penetration testers do)~\cite{miazi_design_2017}.
Threat Hunt teams only search the network for adversaries that existing defenses may have missed~\cite{noauthor_cybersecurity_2017}.
They hunt on the assumption that an adversary has infiltrated the network, when there is no sign of a compromise~\cite{bynum_cyber_2019}.
\textit{\textbf{Unlike the other steps in~\cref{overview}, there is little empirical data about TH team processes, practices, and challenges.}}

\subsubsection{Response (Diagram Step 4)}
\label{response}
The last step in \cref{overview} is \textit{response}. 
Once an adversary is detected, an Incident Response (IR) team is called to respond. 
Incident Response teams are responsible for evicting adversaries that have been found on the network~\cite{nyre-yu_observing_2019}.
IR teams react to adversaries regardless of how they are detected.
IR teams have also been studied by academia and government agencies~\cite{cichonski_computer_2012}.
Researchers have interviewed IR team members~\cite{naseer_real-time_2021,furnell_preparation_2010} and embedded in IR teams~\cite{nyre-yu_observing_2019,sundaramurthy_anthropological_2014}.

\subsection{Threat Hunting Teams and Organizations}
\label{orgs}

In this section we describe how TH is implemented in the private and public sectors.
Although TH is a new discipline~\cite{rasheed_threat_2017,van_os_tahiti_2018}, it is considered an important cybersecurity capability for security practitioners and academia~\cite{milea_hypothesis_2017,gao_enabling_2021}.
TH is mandated for Federal Civilian Executive Branch Agencies~\cite{noauthor_executive_2021} like
  the National Aeronautics and Space Administration (NASA)
  and
  the Department of the Treasury~\cite{noauthor_united_2022}.

Since TH is an emerging discipline of cybersecurity, it is little researched.
The government TH process is particularly important to understand and improve because recent US presidential executive orders~\cite{noauthor_executive_2021} and the US National Cybersecurity Strategy of 2018~\cite{noauthor_national_2018} \textit{mandate} the use of
(third-party)
government hunters on government contractors' networks.

\subsubsection{TH in the Private Sector}

TH can either be done using personnel internal to the organization or using a third party's TH team.
For an internal TH team, organizations either designate SOC personnel to perform hunting or maintain teams dedicated to TH~\cite{lee_who_2016}.
Some companies offer threat hunting as a service, including Booz Allen Hamilton~\cite{noauthor_threat_2022}, Crowdstrike~\cite{noauthor_falcon_2022} and Cisco~\cite{noauthor_talos_2022}.
Most organizations in the private sector opt for internal TH teams~\cite{lee_hunter_2017}, \eg for privacy or intellectual property protection.

\subsubsection{TH in the Public Sector}

The US Government uses both civilian and military teams. 
For example, the Cybersecurity and Infrastructure Security Agency's (CISA) Hunt and Incident Response Team (HIRT)~\cite{noauthor_6_2021} is a team of civilians that performs TH. 
On the military side, all branches of the US military~\cite{symonds_innovating_2017} and some state National Guards~\cite{mcclanahan_169th_2019} have Cyber Protection Teams (CPTs) --- some CPTs conduct TH work along with other cybersecurity functions~\cite{noauthor_cyber_2022}.
Government teams are deployed to federal civilian~\cite{noauthor_executive_2021}, military~\cite{noauthor_u_2015}, and even private sector networks~\cite{noauthor_national_2018,noauthor_6_2021}. 
The Department of Homeland Security includes both CISA and the Coast Guard and thus operates both civilian (CISA) and military (Coast Guard) Hunt Teams. 

\subsubsection{Common Problems Affecting TH Teams}
\label{back_turnover}

Both government and private sector TH teams have substantial personnel turnover~\cite{mciver_closing_2022,oltsik_life_2019,noauthor_how_2020}.
In cybersecurity organizations, public and private, cybersecurity analysts commonly change jobs every two years~\cite{noauthor_cyber_2021}.
Military organizations often rotate personnel every 2--3 years~\cite{bledsoe_how_2023,schmid_military_2022,military_family_advisory_network_effects_nodate}, placing this concern beyond the control of the individual TH teams to address.
Good processes have been shown to mitigate the adverse effects of fewer expert personnel~\cite{nosco_industrial_2020}, helping mitigate the adverse effects of personnel turnover.
Our study describes existing processes as a step toward this goal.

\subsection{TH Frameworks and Processes}
\label{back_frame}

This section presents TH frameworks and processes.
A \textit{TH Framework} is a way of organizing information to assist with the task of hunting.
A \textit{TH Process} is a way of organizing the tasks associated with Threat Hunting across time. 
\ifTECHREPORT
Beyond this material, the appendix gives illustrations of the frameworks (\cref{sec:Appendix-THFrameworks}) and processes (\cref{sec:Appendix-RelatedProcesses}) described here.
\fi

\subsubsection{TH Frameworks}

Both academia~\cite{bynum_cyber_2019} and private sector~\cite{van_os_tahiti_2018} documents suggest three popular frameworks: Lockheed Martin's Kill Chain~\cite{noauthor_gaining_2015}, the Mitre \attack framework~\cite{pennington_getting_2019}, and the Pyramid of Pain~\cite{sqrrl_team_framework_nodate}. 
All three of these frameworks describe adversary activity in a way that assists the defender in categorizing events and focusing their search.
The Kill Chain
\ifTECHREPORT (\cref{fig:KillChain}) \fi
and \attack frameworks outline the steps an attacker takes to carry out a successful attack; \attack goes into greater detail.
The Pyramid of Pain
\ifTECHREPORT (\cref{fig:PyramidOfPain}) \fi
instead assesses what information is most valuable for disrupting adversary activity. 

Researchers often suppose a framework called the \emph{hypothesis method}~\cite{wafula_carve_2019,bynum_cyber_2019,agarwal_cyber_2021,araujo_evidential_2021,horta_neto_cyber_2020}.
The hypothesis method is when the TH team outlines a possible intrusion that could have occurred on the network and then tests that hypothesis using the data available.
An example hypothesis is: \textit{``Advanced Persistent Threat \#0 exploited CVE-2021-44228 to compromise a VPN Server then moved laterally to the domain controller.''}  

An alternative to hypothesis hunting is \textit{data-driven unstructured hunting}~\cite{security_threat_2022,security_cyber_2022,van_os_tahiti_2018}.
Here, threat hunters search for adversaries without a hypothesis, guided by statistical or behavioral analysis to identify adversary activity.  
Some authors consider this as hunting without a process~\cite{security_threat_2022}.
Others say it is less efficient than the hypothesis method~\cite{noauthor_attack_2021,security_cyber_2022}.

These frameworks and methods are complementary~\cite{van_os_tahiti_2018}.
For example, a TH team could build a hypothesis using the steps in the Kill Chain framework. 
The team could then look up the associated techniques using the \attack framework and prioritize the search based on the Pyramid of Pain.

\subsubsection{TH Processes}


Currently, the most common way to hunt for an adversary is \adhoc, \ie without a formal process~\cite{lee_hunter_2017}.
Some TH teams may have privately documented processes, but public descriptions are rare.
We describe what is known.

\myparagraph{Private sector processes:}
To address the lack of standardized processes and definitions, four organizations from the Dutch financial sector shared a process known as the Targeted Hunting integrating Threat Intelligence (\tahiti) process~\cite{van_os_tahiti_2018}. 
\tahiti describes the behavior of internal threat hunters rather than third-party/external hunters, and thus makes assumptions about the network information that the threat hunters have.
\ifTECHREPORT
The \tahiti process is depicted in \cref{fig:TahitiProcessModel}.
It provides a process for structured, hypothesis-based TH. 
\fi

The private sector has made other attempts to explain TH processes but none as detailed as \tahiti.
Endgame published \textit{The Endgame Guide to Threat Hunting}~\cite{ewing_endgame_nodate,scarfone_hunters_nodate}.
They proposed a 4-phase process with 6 hunting steps. 
Other TH guides did not provide a TH process~\cite{noauthor_guide_2020}. 

These TH process models --- \tahiti and Endgame --- are prescriptive, not descriptive.
\tahiti was derived from a private sector round table, describing what should occur, not necessarily what does occur. 
Endgame is similar.

\myparagraph{US government processes:}
There is no public documentation of the TH process used by Government TH teams. 
The closest work is by Trent \etal~\cite{trent_modelling_2019}.
They created a \textit{cognitive} model of activities performed by US Army CPTs. 
Their goal was not specifically to outline a Threat Hunting \textit{process}, but as CPTs perform TH their findings are somewhat relevant.
\ifTECHREPORT
The cognitive model found by Trent \etal is in~\cref{fig:TrentCPTWorkModel}.
\else
\fi

\myparagraph{Academic perspectives:}
Previous academic works on TH methodologies do not describe current TH practices.
They propose novel approaches to assist TH teams~\cite{agarwal_cyber_2021,wafula_carve_2019}.

\myparagraph{Comparing private sector and govt. TH processes:}
Comparison between the TH processes proposed by \tahiti~\cite{van_os_tahiti_2018} and Trent \etal~\cite{trent_modelling_2019} is imperfect for two reasons:
(1) \tahiti was focused on TH while Trent \etal covered TH as well as other cybersecurity roles; and 
and
(2) \tahiti is a process while Trent \etal is a cognitive model.
Nevertheless, our analysis of these works suggests differences between private sector and government TH teams. 
\tahiti is for internal TH teams, while Trent focuses on third-party CPTs. 
This means that \tahiti assumes more shared context across missions, \eg keeping ``investigation abstracts'' in a backlog for future hunts. 
In contrast, the external hunts from Trent are more self-contained, with tasks devoted to ``Planning and Logistics'' and ``Closure''. 
Perhaps related, we also observe that \tahiti is more abstract while the Trent model is more detailed, including tasks such as Forensic analysis as distinct from Host, Malware, and Network analysis.
Finally, the \tahiti TH process uses the hypothesis method, while Trent \etal does not indicate a hypothesis (even though generating a hypothesis would be a unique cognitive task). 

\subsection{Summary and Unknowns} \label{sec:Background-Unknowns}

Although threat hunting is a mandated government function, little is known about TH team processes. 
The most detailed TH process is \tahiti, a process described by threat hunters from four private sector institutions~\cite{van_os_tahiti_2018}.
However, government TH processes likely differ from private sector ones
because government teams often hunt on third-party networks rather than within their own networks~\cite{noauthor_national_2018}.
The TH process used by government teams is unknown. 

Since most organizations that perform TH do so without a formal process, understanding TH processes outside of the ``internal private sector team'' context (\tahiti) will help organizations developing their own TH processes. 
Typical TH team practices and challenges are also undocumented --- learning these would benefit all TH organizations, whether or not they that have a TH process. 
One specific area of interest is the effect of turnover and integration of newcomers, a concern shared by private and public-sector TH teams.


\section{Research Questions and Methodology}

Historically, the US government have driven the creation of cybersecurity standards~\cite{spring_review_2021}.
For example, after the 2015 Office of Personnel Management (OPM) data breach~\cite{jason_chaffetz_opm_nodate}, many private organizations sought to improve their processes from shortcomings observed in the OPM process~\cite{noauthor_lessons_2015,lemos_5_2016,noauthor_lessons_2016,kerner_lessons_2017}.
Describing the government TH process may be beneficial for the same reasons, both as an example and to assist in identifying shortcomings. 
The government is among the sectors most often targeted by cyberattackers~\cite{singleton_x-force_2022}.
The US government's cybersecurity defenses, including its TH processes, are thus of great interest.  
We provide a first view.

\subsection{Research Questions}
Describing all US government TH processes is beyond the scope of a single study.
As a step toward this goal, in this work we examine TH processes from one branch of the government, the US Department of Homeland Security (DHS).
In this context, we address two research questions:

\begin{itemize}[itemsep=0pt,leftmargin=*]
 \item \textbf{RQ1:} What processes are used by DHS TH teams?
 \item \textbf{RQ2:} What shortcomings exist with current DHS TH processes and what might be done to alleviate them?
\end{itemize}

\subsection{Statement of Positionality}
\label{statement}
One of the authors is a former DHS Threat Hunter. 
We acknowledge that this author's background shaped, both directly and indirectly, our study design, recruiting, analysis, and findings~\cite{dodgson2019reflexivity}.
For example,
 their professional experiences informed our interview protocol,
 and
 we recruited their former colleagues as subjects.
The researcher's relationship with some of the subjects may have had positive and negative effects:  
  responses could be biased by the relationship, but
  they may also be enriched because interviewer-subject trust had already been developed.
Despite the potential biases, we emphasize that access to TH personnel has been a barrier to research.
Capturing aspects of DHS TH experiences is valuable, even if incomplete. 

\subsection{Study Design}
\label{design}

An Interview methodology was chosen for this study as a result of the data available and the population being studied. 
We felt the prior work of Trent \etal~\cite{trent_modelling_2019} and van Os \etal~\cite{van_os_tahiti_2018} provided enough initial structure to frame the collection instrument and initial analysis. 
Long-form ($\sim$1hr) interviews allowed us to make the most of a relatively small sample size.  
Little information exists on the processes used by government Threat Hunting teams, so our study will direct future research. 

We considered and decided against other methodologies.
It would be difficult to capture the complexity of the TH process through a survey, which would not allow for iteration on the collection instrument to pick up unexpected intricacies~\cite{wilson_evaluation_2015}.
Additionally, since the population of government threat hunters is small and difficult to access, a meaningful sample size for a survey would be difficult to achieve~\cite{bartlett_organizational_nodate}. 
We also considered a grounded theory methodology, which is suitable when no prior theory or framework exists~\cite{klaas-jan_stol_grounded_2015}.
In our case, however, we had access to prior work done by private TH teams (\tahiti) and studies on TH-adjacent cybersecurity teams (Trent et al.) which we used to guide our study. 

The TH teams we studied have three roles:
\textit{Leadership}, \eg officers, deal with TH strategic concerns but do not deploy with the team.
\emph{Team Leads} deploy with a hunt team and act as the on-site manager. 
\emph{Analysts} deploy and perform the analytic tasks associated with hunting. 
Each role may have different perspectives on the TH process, challenges, and solutions, so we recruited subjects in each role. 

\subsection{Recruitment and Subject Demographics}
Government Threat Hunting teams are a difficult group of practitioners to study due to the small size of the teams and the sensitive information they often deal with. 
One author's US government security clearance and previous TH duties allowed us access to these TH team members and helped ensure no sensitive information was disclosed.

Participants were recruited for interviews through the authors' professional network.
Emails were sent to TH members with varying experience and roles. 
\new{At both organizations, subjects were representative of multiple internal teams and analyst populations, including primary organizational divisions.}
The response rate from direct contacts was 59\% (10/17).
One additional subject offered to be interviewed after hearing about the study. 
6 out of the 11 participants had previously worked with one author as a peer (1), subordinates (3), or managers (2).
Participants were uncompensated volunteers \new{to avoid conflict of interest with their colleague (the interviewer).}

\new{We tried to go beyond our professional network, but recruitment was challenging. We contacted TH analysts from other agencies (DOE and DOD) and received one response. We ultimately excluded the other organizations from the study and focused on the one to which we had access.}

\myparagraph{Subject Demographics:}
\label{subjects}
The distribution of subjects by organization and by position is in~\cref{table:SubjectDemographics}.\footnote{\ifTECHREPORT\cref{subs_vs_exp_mis} shows \fi subjects' experience in years and \# of missions.}
Some subjects had operated on teams in multiple organizations within the last three years so the count in~\cref{table:SubjectDemographics} exceeds 11. 
As precise titles could de-anonymize subjects, subjects are mapped to three general job roles: Leadership, Team Lead, and Analyst.

\begin{table}[]
\caption{
    Subject breakdown by organization and position.
    }
    \begin{small}
\parbox{.45\linewidth}{
\centering
\renewcommand{\arraystretch}{0.85}
\begin{tabular}{cc}
\toprule
\textbf{Organizations} & \textbf{\# Subjs.} \\ \midrule
DHS Organization \# 1 & 10 \\
DHS Organization \# 2 & 3 \\
\bottomrule
\end{tabular}
}
\hfill
\parbox{.38\linewidth}{
\renewcommand{\arraystretch}{0.85}
\begin{tabular}{cc}
\toprule
\textbf{Position} & \textbf{\# Subjs.} \\ \midrule
Leadership & 4 \\
Team leads & 4 \\
Analysts & 3 \\ \bottomrule
\end{tabular}
}
\label{table:SubjectDemographics}
\end{small}
\end{table}

\subsection{Interview Procedure}

\myparagraph{Interview protocol creation:}
The main body of our instrument focused on the knowledge gaps identified in~\cref{sec:Background-Unknowns}: eliciting the TH process and understanding its challenges.
Guided by previous TH literature, we asked about analysis frameworks and process automation~\cite{bynum_cyber_2019,agarwal_cyber_2021,van_os_tahiti_2018}.
Based on the frequency of personnel turnover, we also had a line of questions about incorporating new members.

\myparagraph{Interview protocol refinement:}
Following best practices~\cite{chenail_interviewing_2011}, we conducted two mock interviews.
The primary researcher (who has TH experience) was interviewed by the other researchers. 
The protocol was tuned for clarity.
Transcripts from these mock interviews were not used in analysis.

After the mock interviews, we divided the real interviews into two stages. 
In the first stage, we held two interviews to pilot our protocol, one with a CGCYBER team lead and one with a CISA team lead. 
After these interviews we reviewed the transcripts, assessed validity, and made changes as needed.
Following this pilot, we added 2 questions and 3 follow-up questions, and re-worded one follow-up question.
Over the course of the entire study, 92\% of the main questions (22/24) were held constant.
As there was little change in the instrument, our results include the 2 pilot interviews.
\ifTECHREPORT
Further details are in~\cref{protocol}.
\fi

\ifTECHREPORT
The final semi-structured interview protocol is summarized in~\cref{table:InterviewProtocolSummary} and detailed in~\cref{sec:Appendix-InterviewProtocol}.
\fi
All interviews were conducted by one author, who had security clearance (for national security,~\cref{ethics}) and prior TH experience (so that they could ask domain-appropriate follow-up questions).
Interviews lasted 1 hour and used Microsoft Teams. 

{
\renewcommand{\arraystretch}{0.1}
\begin{table}
    \centering
    \small
    \caption{Summary of interview protocol. The first column is the topic and the number of questions for that topic.}
    \label{table:InterviewProtocolSummary}
    \begin{tabularx}{\columnwidth}{>{\centering\arraybackslash}m{2.5cm} X}
        \toprule
        \multicolumn{1}{c}{\textbf{Topic (\#)}} & \multicolumn{1}{c}{\textbf{Example questions}} \\
        \midrule
        Demographics (2) & 
        \begin{itemize}[topsep=0pt, partopsep=0pt, leftmargin=*, labelsep=2mm, nosep, before=\vspace{-0.6em},after=\vspace{-0.6em}]
            \item How many missions have you been on?
            \item How many found an adversary?
        \end{itemize} \\
        TH process (11) & 
        \begin{itemize}[topsep=0pt, partopsep=0pt, leftmargin=*, labelsep=2mm, nosep, before=\vspace{-0.6em},after=\vspace{-0.6em}]
            \item Draw your team's TH process.
            \item What parts are problematic?
            \item In what ways do you incorporate hypotheses into the process?
            \item (Critical incidents) Tell us about 1-2 missions where the process failed?
        \end{itemize} \\
        New members (6) & 
        \begin{itemize}[topsep=0pt, partopsep=0pt, leftmargin=*, labelsep=2mm, nosep, before=\vspace{-0.6em},after=\vspace{-0.6em}]
            \item How long does it take a new team member to become productive?
            \item What is a good measure of a member's expertise?
        \end{itemize} \\
        \bottomrule
    \end{tabularx}
\end{table}
}


\subsection{Ethics and National Security}
\label{ethics}

This study was approved by our institution's Internal Review Board (IRB). 
A signed consent form was collected from participants before their interviews. 

The studied organizations have small TH communities so care was taken to anonymize subjects. 
Specifically, we removed identifying information from quotes, and we describe and quote subjects in terms of generic job roles (\cref{table:SubjectDemographics}).

Due to the sensitive information being discussed, we took measures to ensure that no classified information nor data from customer sites was collected. 
The interviewer reminded each subject of the unclassified nature of the research. 
All audio was reviewed for sensitive information before being sent to the transcription service. 
The transcript was then reviewed again by the research team before being used for analysis.

\subsection{Data Analysis}
\label{data_analysis}

We used three types of analysis on the resulting transcripts.
  Process coding and
  inductive process discovery were used for RQ1.
  Thematic coding was used for RQ1 and RQ2.
\ifTECHREPORT
All codebooks are in~\cref{sec:Appendix-Codebooks}. 
\fi

\subsubsection{RQ1: Process Identification} \label{sec:Method-DataAnalysis-RQ1}

First, process coding was done as described in Salda\~na~\cite{saldana_coding_2021}.
In this method, a pre-existing process is represented in a codebook used to analyze (code) a transcript.
We created a codebook from \tahiti using \ifTECHREPORT \cref{fig:TahitiProcessModel} and\else \fi their descriptions of hunt triggers.
We likewise created a codebook for the Trent model. 
The subjects' descriptions of their TH process were then coded against these process codebooks to assess fit. 
\footnote{The TH process description was the first part of the interview. The \tahiti and Trent codebooks were only used on this part of the interview. It was rare for a subject to mention a new component of their process unprompted after their process description.} 

In our results (\cref{rq1}), neither \tahiti nor Trent was a good fit, so we induced a TH process model from the interview data.
First, all nodes from all subject diagrams were placed into one interconnected diagram. 
Similar nodes were combined.
When possible, more precise nodes took precedence over general nodes. 
If a node was only on one diagram and not mentioned in multiple subjects' interviews, that node was removed.

\new{As this process was relatively objective, a single analyst combined nodes and made the diagram. Another analyst iteratively reviewed the resulting process model.}


\subsubsection{RQ2: Shortcoming Identification} \label{sec:Method-DataAnalysis-RQ2}

\new{Following process coding and diagram discovery, further analysis was conducted on the transcripts. Since the interview protocol contained primarily focused questions, the interviews proceeded in a readable linear fashion. This allowed one analyst to re-code the transcripts} using thematic coding as described in Guest \etal \cite{guest_applied_2012}.
Memos were written by the primary analyst. 
\new{This researcher} arranged the 636 memos into themes.
Codes were iteratively refined, ultimately yielding 57 codes under 10 topics\ifTECHREPORT \ (\cref{table:FullThematicCodebook}). \fi 
\new{A second analyst assessed reliability, using the codebook and excerpts from 6 of the 11 transcripts. For completeness, excerpts were selected from each set of Question-Answer in the interview protocol. The measured agreement was fairly high (Cohen's $\kappa=0.82$).}



\myparagraph{Completeness of results:}
Saturation was measured after all 11 interviews were complete.
We measured saturation following Guest \etal \cite{guest_how_2006}, by measuring the number of (cumulative) new codes appearing in each interview. 
\ifTECHREPORT
The number of cumulative codes observed in each interview and in the previous interviews, charted in \cref{sat_fig} indicates that saturation was achieved after seven subjects, with no new codes being observed in the last five interviews. 
\fi
All organizations and all position categories had been represented at that point, indicating substantial homogeneity. 
Codes on a per-interview basis were also charted.
\ifTECHREPORT
We found that each interview covered many topics ($\geq$30 codes per interview, see \cref{sec:Appendix-SaturationCharts}).
\fi


After analysis was complete we performed member checking~\cite{birt_member_2016} by circulating our results to two experienced subjects (one team lead and one member of leadership). 
They felt properly anonymized and that their TH data was represented.


\subsection{Limitations and Threats to Validity}
\label{limitations}
Like many qualitative studies, our primary limitations are the sample size (11 subjects) and the reliance on self-report.
\begin{itemize}[itemsep=0pt,leftmargin=*]
    \item \textit{N=11}: 
  Guest \etal argue as few as six interviews can suffice if the population is homogeneous and the data collected is specific~\cite{guest_how_2006}.
We believe this is the case here. 
Samples of $\sim$10 are common in interview studies~\cite{huck_wake_2022,pettigrew_making_2012,de_gramatica_it_2015,sahin2023investigating}.
Also note our sample is relatively large compared to the population, which is $\sim$5,000 at the studied organizations.
\item \textit{Self-report:} 
For validity, we followed best practices in interview instrument creation~\cite{chenail_interviewing_2011} including conducting 2 internal mock interviews and slightly modifying the instrument after the first 2 subject interviews.
We performed member checking~\cite{birt_member_2016} by circulating our results to two experienced subjects (one team lead and one member of leadership). 
They both felt properly represented.
Unfortunately, triangulating against public documents was not possible, as those from TH teams lack process information~\cite{cybersecurity_and_infastructure_security_agency_malicious_2022} and internal documents are confidential.
\end{itemize}



We note two additional threats.
First, the analysis was done primarily by one researcher. 
Bias is mitigated by measurements of inter-rater agreement on a subset of the data.
Second, our study may not generalize, \eg to the rest of the government or to private sector teams.
Government teams and agencies differ by mission, training, etc.
The differences between the \tahiti process and the Trent model suggest differences between public and private sector teams. 

\section{Results}

\subsection{RQ1: TH Processes of DHS Teams?} 
\label{rq1}

\begin{tcolorbox}
The process described by Government practitioners differed from the \tahiti and Trent processes. 
We induce a unified process model capturing the participants' TH processes, comprising 7 stages and 25 distinct activities (\cref{fig:CoarseInducedDiagram}).
\end{tcolorbox}

Each subject provided two types of data to indicate the process they used. 
First, at the beginning of their interview, 10 out of 11 subjects provided a process diagram.
(The remaining subject did not feel sufficiently familiar with the process.) 
Second, in the remainder of the interview they described their team's process. 
We used process coding to compare our subjects' TH processes  to the \tahiti~\cite{van_os_tahiti_2018} and Trent~\cite{trent_modelling_2019} models. 
Our observed TH processes did not match (\cref{process_coding_rq1}), and we describe the induced model in~\cref{observed_rq1}.


\subsubsection{TH Process Comparison to Related Work}
\label{process_coding_rq1}

We coded processes against the \tahiti and Trent models. 

\myparagraph{\tahiti:}
We found \tahiti a poor match to our subjects' processes.
88\% of the \tahiti codes matched to activities described by the subjects.
However, every subject described at least one activity that occurred in addition to the \tahiti process, including Baselining (8 subjects), Sensor Placement (8 subjects), Team Arrival (5 subjects), and Customer Meetings (5 subjects).
\ifTECHREPORT
Note that all these activities are included in \cref{fig:DetailedInducedDiagram}.
\fi
We suggest two reasons for this poor fit:

\mysubparagraph{(1) Internal Team Assumed:} 
Some tasks that are only required for external organizations were omitted by \tahiti. 
For example, when describing their process 8 subjects were mentioned placing sensors on site prior to the team's arrival. 
\tahiti discusses data sources and ensuring data availability, but its concerns are different than our subjects' concerns about sensor placement.
Sensor placement is not discussed by \tahiti but was important to our subjects because their organizations operate on external third-party networks. 

\mysubparagraph{(2) Only Hypothesis Hunting Described:}
Some tasks described by subjects are more important for data-driven hunting than hypothesis-driven hunting. 
For example, baselining was described by subjects as a 1-3 day process in which the TH team, deployed on an unfamiliar network, takes time to document typical network behaviors. 
Subjects indicated that baselining was especially important for filtering out false positives and for behavior analysis. 
However, since \tahiti describes hypothesis-based hunting, it does not cover data-driven hunting techniques like behavior analysis. 

\myparagraph{Trent:}
We found Trent a poor match to our subjects' processes because of the unit of analysis.
%
The Trent model was designed to distinguish between cognitive tasks.
Our subjects more often drew administrative or temporal distinctions between tasks. 
For example, the Trent model describes four types of analysis: Network, Host, Malware, and Forensic.
Our subjects instead described their analysis work in terms of two processes, a manual behavior-driven loop and an automatic alert-driven loop.
Both loops involved host and network analysis --- although these are distinct cognitive tasks, most
subjects did not perceive them as distinct process components. 
We observed the same dynamic with baselining --- Trent groups baselining activities with similar activities in other phases of a hunt, while our subjects emphasized the importance of a separate task for baselining distinct from the analysis step of the process.
Numerically, 21\% of the process components mentioned by our subjects did not match any of Trent code.

\myparagraph{Incorporation of Different Frameworks:}
Each subject was specifically asked about their use of the hypothesis method and three other models observed in TH descriptions across academia and the private sector (see \cref{back_frame}).  
\ifTECHREPORT
The results can be seen in \cref{framework_table}. 
\fi
In both organizations, Mitre's \attack framework was the most used, although its role differed across subjects. 

\subsubsection{Description of the Induced TH Process}  
\label{observed_rq1}

\label{rq1_walkthrough}



\begin{figure*}[ht]
\includegraphics[width=.93\textwidth,height=6cm]{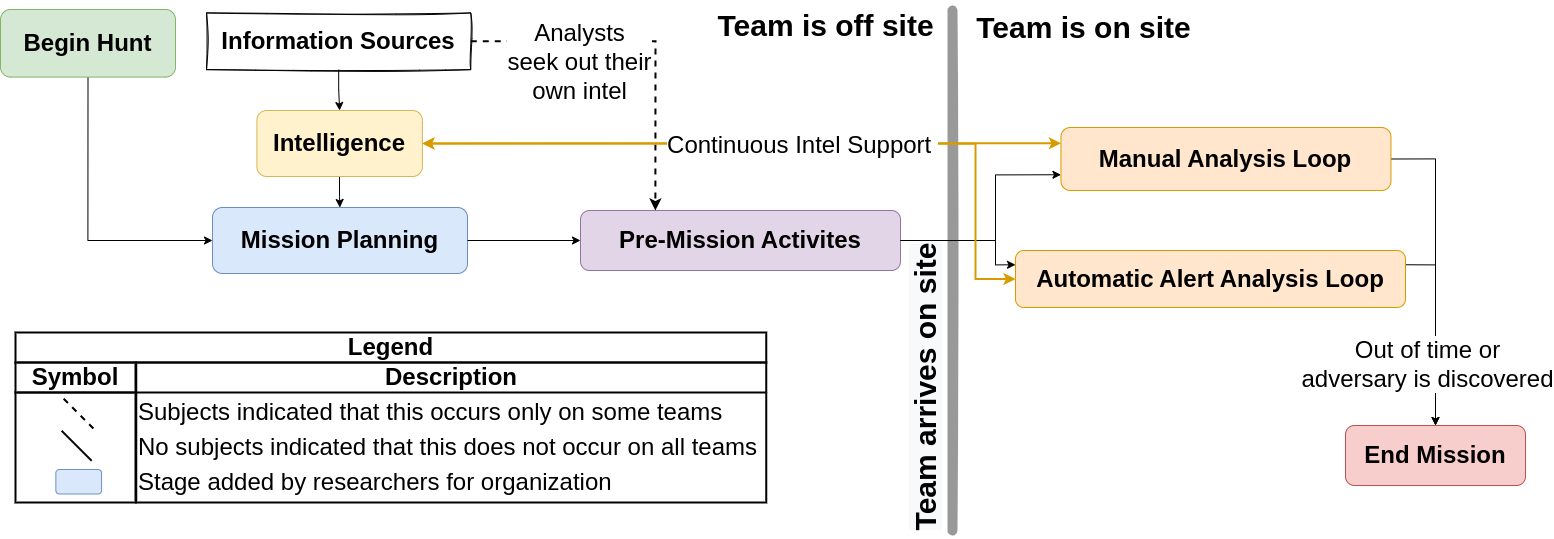}
\caption{
  Unified TH process induced from interview data. 
  \ifTECHREPORT
  A detailed version is given in~\cref{fig:DetailedInducedDiagram}.
  \fi
  }
\label{fig:CoarseInducedDiagram}
\end{figure*}

We concluded that our subjects' TH processes differed from the available literature.
We induced a unified process model following the method given in~\cref{sec:Method-DataAnalysis-RQ1}. 
\cref{fig:CoarseInducedDiagram} shows the high-level result, with activities grouped into seven stages based on our judgment.
\ifTECHREPORT
A detailed version is in the appendix (\cref{fig:DetailedInducedDiagram}).
\fi
This section walks through~\cref{fig:CoarseInducedDiagram}.

\myparagraph{(1) Begin Hunt:}
Subjects described two ways to begin a TH mission.\footnote{Subjects used different names for an individual hunting operation. We use the term \textit{mission}. Third-party TH teams consider a mission as one engagement with a specific customer over a few weeks. In contrast, internal TH teams hunt continuously, with few-day missions by hypothesis.}
(1) In a \textit{proactive} mission, the customer has no suspicion of adversarial activity but still wants a TH team to check. 
One subject said: \inlinequote{Some of our [missions]...people just say: `Well, we're interested in having you come see if anyone's here'.} 
(2) In a \textit{triggered} mission, there is no specific indication of compromise but the customer believes an adversary may be undetected on their network and requests a TH team.
In one subject's words: \inlinequote{Maybe 
you get...intelligence that says, `...a system that belongs to you may be communicating to a malicious command and control infrastructure'.} 

\myparagraph{(2) Mission Planning:}
In the mission planning task, the TH team coordinates with the customer. 
A leader subject said: \inlinequote{I think [planning is] the most critical piece...This is really what I think creates a useful engagement.}
This process typically starts with a customer's request, then a survey of their system and network admins, and then TH-customer meetings (\eg \inlinequote{multiple technical phone calls...to discuss any details}). 
Subjects described three components of mission planning: scoping, hypothesis creation, and mission plan creation. 

\textit{Scoping} is when the TH team decides on what parts of the network will be included in the hunt mission. 
Organizations may constrain the hunt in ways such as
  parts of their network,
  types of threats considered,
  and
  time limit.
In one subject's words: \inlinequote{We got to figure out what kind of environment we're going to be working in, how many endpoints, how many users, inventory, ...how much data...
stuff like that. We have a scoping questionnaire we send to the partner.}
Internal hunt teams often have a fixed scope of operation. 

During \textit{hypothesis creation}, a TH team generates a hypothesis to test in the hunt.
Some teams do this during planning; others create hypotheses after deployment; others do not use hypotheses.
Of the three studied organizations, two formally document hypotheses in the mission plan or elsewhere. 

The \textit{mission plan} documents the structure of the mission.
It could include a timeline with associated deliverables, specific TH objectives, and 1-3 hypotheses if any.  
Not every team created a mission plan document. 
It may be unnecessary for internal teams as their hunt missions tend to be shorter. 

Subjects in leadership positions often mentioned the importance of objectives in the mission plan, 
\eg
\inlinequote{The really important thing for TH is you go in with specific objectives. You're not just trying to find all bad activity.} 


Plans and objectives are not always communicated to the whole team.
When asked if their team used a mission plan, one experienced analyst said: 
\inlinequote{Yes, but I didn't have access ---
I was just a lowly [low rank] at the time. It was the team lead who ... would come up with it.}

\myparagraph{(3) Collect Intelligence:}
Subjects described 3 intelligence collection tasks. 
(1) When a specific trigger exists, intelligence is tailored around that trigger.
(2) On proactive missions, current attacker trends are used: 
  \inlinequote{If there's a current prevalent exploit that's being used, like that you're seeing in the news...
  we're gonna go ahead and look for those.}
(3) Regardless if the hunt is proactive or triggered, teams will often collect and upload additional Indicators of Compromise (IOCs) to achieve increased coverage: 
  \inlinequote{we'll...load IOCs from all the threat intelligence providers.}

Subjects provided many example intelligence sources but 4 sources were recurring:
(1) A distinct intelligence team supporting the TH team;
(2) Publicly Available Information (PAI) like social media or news feeds;
(3) Open-Source Repositories like InQuest's Indicators of Compromise (IOC) database~\cite{noauthor_inquest_2023};
and
(4) Classified or Subscription Feeds.

\myparagraph{(4) Pre-Mission Activities:}
Before a TH team begins a mission, they fine-tune their tools using the Intelligence and the mission scope, hypothesis, and plan.
%
One subject described typical pre-mission activities and how these can vary between internal and external hunts:

\blockquote{So you plan the mission...dates [and] goals...
And then you'll push that into developing your specific tools and analytics.
If you're hunting yourself [\ie internally], you probably already have most of your sensors and tools in place 
...
If you don't have tools or specific analytics to cover what you're going after, then you need to develop or buy or get those plugged in.}

\myparagraph{(5/6) Manual and Automatic Analysis Loops:}
At this point, the TH team deploys to the customer's site and spends 1--4 weeks hunting.
Most of the TH time is spent here.

Subjects described two different modes of analysis.
Both modes included cyclical tasks so we term them ``loops''. 
In the Manual Analysis Loop, TH team members enrich their baseline and perform manual analysis of the collected data, looking for potentially malicious behaviors. 
In the Automatic Alert Analysis Loop, members triage sensor notifications of possible IOCs. 
These loops continue concurrently and members are assigned to them at the team lead's discretion.
A team lead summarized: \inlinequote{Those two processes...are our constant back and forth over a couple weeks that we're doing on-site until we can either find something or not.}


One subject described these loops (``paths'')\ifTECHREPORT, cf.~\cref{fig:DetailedInducedDiagram}\fi: 
\inlinequote{So the first path is your alerting path [\underline{Note: the automated loop}]. Those are your indicators and signature based detections ... And then the second path [\underline{Note: the manual loop}] I see kind of starts with understanding the environment which is some baseline analysis...that path then feeds into your two like core detections, which are ... your behavioral analysis ... and then you also have anomaly based analysis.}

\myparagraph{(7) End Mission:}
The goal of a hunt mission is to detect adversaries, if they exist on the network. 
If adversary activity is observed, the hunt is over and a response is necessary. 
One subject said: \inlinequote{If we find commodity malware on the computer, it may not require a lot of resources...If it's something bigger...we're probably going to shift into IR mode.}


At the end of the mission, TH teams report their findings and provide recommendations:
\inlinequote{We come back to those hypothesis in our final report, too....[the report] goes to the entity and back to our leadership saying, `We tried to see if your exchange server was compromised by doing X, Y, and Z. Here's some things we investigated as a part of that. Here's our conclusion or what that led us to believe.'} 


\subsubsection{Process management}
\label{management_rq1}

\myparagraph{Process Creation:}
\label{creation_rq1}
Process creation differed across organizations. 
One organization's process was devised by a group of team leads.
At another organization, the TH process was created by 1-2 people in leadership positions and was deliberately imitative of other government TH teams. 
This imitation lets their TH teams interoperate with other organizations, enabling personnel sharing in the event of a national cyberattack.

\myparagraph{Process Changes:}
\label{changes_rq1}
At both organizations, the TH process changes frequently, and changes can be proposed by any experienced personnel: \inlinequote{everyone who had prior experiences or is now part of the [team] and been on a mission...can submit edits.}
Subjects describe the process changes positively, in terms of continuous improvement --- 
  \inlinequote{We're constantly looking for how we can improve... So basically after every mission... we do a hot wash and we say: `Okay, what could have gone better?'}.
One team lead did warn that process change in the middle of a mission can be problematic: \inlinequote{If you change the tactics too often, you are going to tire out your analysts ... You're gonna tire out yourself and you're gonna get confused when you go to write the final report.} 
They implement changes between missions instead.

\subsection{RQ2: Process shortcomings and solutions?} 
\label{rq2}

\begin{tcolorbox}
\cref{rq2_table} shows process issues, proposed solutions, and open questions. 
Common process challenges are (1) determining expertise of team members (to assign appropriate tasks); and (2) Improving automation.
\end{tcolorbox}

This section describes TH process challenges noted by $\geq3$ subjects across both organizations, and possible solutions where available.
Shortcomings are ordered by frequency. 
To avoid biasing to a single subject's ``pet peeves'', in this section all subjects are quoted at least twice and no more than \maxquote times. 

\begin{table*}[]
\small
\caption{
    Process challenges and proposed solutions noted by subjects.
    Open problems have no consensus solution or occur when subjects verbalized difficulty or uncertainty. 
    Only challenges noted by $\geq3$ subjects and observed in both orgs. are included.
    }
\renewcommand{\arraystretch}{0.83}
\begin{tabularx}{\textwidth}{XcXX}
\toprule
\textbf{Observed process challenge} & \textbf{\# subjects (\# orgs)} & \textbf{Proposed Solutions} & \textbf{Open Problems} \\
\toprule
Poor measures of expertise (\cref{expertise_rq2}) & 6 (2) & Make training more tool oriented & Determining TH expertise? \\ \midrule
Insufficient automation (\cref{automation_rq2}) & 6 (2) & Automate baselining \& set-up tasks & Will automation hinder the analysts? \\ \midrule 
Teams lack needed data (\cref{data_rq2}) & 5 (2) & Deploy sensors ahead of time & What data to collect? \\ \midrule
Inappropriate process detail (\cref{detail_rq2}) & 4 (2) & Give team leads flexibility & \textit{None mentioned by subjects}\\ 

High turnover (\cref{turnover_rq2}) & 3 (2) & Pair new and experienced members; &  \textit{None mentioned by subjects}\\ 
 & & Enhance process documentation & \\\bottomrule
\end{tabularx}
\label{rq2_table}
\end{table*}

\subsubsection{Identifying expertise}
\label{expertise_rq2}

\myparagraph{Problems:} 
TH team leaders want to assign members different tasks depending on the members' expertise.
Many indicators of expertise exist --- three common indicators are external certifications, internal certifications, and training.
Subjects seemed to dislike these. 
Instead, they used \textit{ad hoc} indicators such as experience and time spent working off-hours.

\mysubparagraph{(1) Certifications}
Six subjects did not believe certifications were a good measure of expertise.
One subject 
was skeptical that certifications measured even baseline knowledge: \inlinequote{It's really hard to define what a baseline of cybersecurity understanding is ... You can't really say it's having a certain certification because there's plenty of people that have certifications that don't know what they're talking about.}

\mysubparagraph{(2) Internal Certification}
One organization maintains an internal certification as mentioned by five
subjects.
One subject stated that it helped the team be interoperable with other government TH teams.
Three subjects voiced dissatisfaction with the certification stating it was irrelevant or poorly implemented: \inlinequote{
I have some mixed feelings on [the certification]...It doesn't 
...apply to us all that well and the proficiency needed to complete it isn't
...much either.} 

\mysubparagraph{(3) Training}
Two subjects, both inexperienced, spoke favorably of training as an indicator of expertise: \inlinequote{I think it's just training experience, time with the tools, time with the knowledge base --- I think training has a big part of it though. And I think just hands-on is pretty key as well, which the training could help with.
}
Two experienced subjects were more cautious. For example, one said: \inlinequote{There is a mild or weak correlation between number of training courses and analyst success}.
Four subjects spoke unfavorably of training as an indicator of expertise. 
A team lead said \inlinequote{having a [certain] course under your belt or any other course, usually doesn't give me immediate confidence.}.
One analyst strongly critiqued their required training courses:
\inlinequote{We have [an agency] requirement...
just a massive waste of time. \textless laugh\textgreater} 
However, this analyst recognized that
some trainings can help: 
\inlinequote{I've given the team several trainings [about tools]...little
exercises to help people get up to speed.} 

\myparagraph{Solutions:}
Developing indicators of expertise remains an open question.  
Subjects had some suggestions, which we list next, though each idea was opposed by $\geq$1 subject.

\mysubparagraph{(1) Time with tools}
The importance of time with the tools was a theme that often accompanied discussions about training, as an indicator of expertise or a tactic to integrate newer members. 
When subjects discussed the value of training, they often spoke in terms of whether it let them improve their skills with their tools. 
For example, one subject said: \inlinequote{looking at the [training]...
it's a good experience
but...
they don't use the tools we currently use...It doesn't provide the training that we need per se, for the tools that we use.}




\mysubparagraph{(2) Working on personal time}
Working off-hours was viewed more positively than training or certifications.
Five subjects said doing cybersecurity activities on personal time was a good metric for potential or expertise. 
Subjects made comments like: 
\inlinequote{I can't really put my finger on a single thing that...explains our expertise. I would say, a big part of it is just being willing to kind of play around with the tools and even in their own time. I mean, we have a lot of downtime between missions where we do trainings, but I'm thinking more, even within that time-frame, they have a lot of free time where they can kind of play around with the tools or do you know, capture the flags
...that almost has been a bigger indicator.}
We note that not all analysts will be able to devote free time and that making this an indicator of expertise could have undesired effects such as marginalizing certain otherwise high performing analysts as has happened in similar cyber security fields\cite{fulton_vulnerability_2023}.

\mysubparagraph{(3) Experience}
Ten subjects mentioned experience (\eg the number of missions) in the context of expertise. 
Five subjects were generally positive about experience as an indicator of expertise, while four demurred.
One subject said:
\inlinequote{I think experience is a key metric.
...most of the guys that we got that had already done threat hunting [before joining our organization] are our best analysts}. 
A common counter-argument was: 
\inlinequote{Experience does not equal quality...just because someone has experience doesn't mean they're good at their job.}

Subjects who affirmed experience as an indicator of expertise often did so hesitantly, recognizing it as a heuristic. 
For example, the positive subject quoted earlier also said 
\inlinequote{You...have people who've been on a hundred missions and they just don't have the aptitude or the thought ability. You...have people who have never been on a mission who have the aptitude and who are gonna outperform people who have been on 20 missions. ... I think it's a little hard to answer that question directly, but I would say the biggest \textless hesitates\textgreater\ there is a correlation between number of missions and analyst success.}

\subsubsection{Improving Automation}
\label{automation_rq2}

\myparagraph{Problems:}
Automation is often emphasized as a way to reduce costs or to discover adversaries more quickly.
The subjects indicated their automation was insufficient or ineffective. 

\mysubparagraph{(1) Insufficient Automation}
One issue was that both organizations had little automation. 
When subjects were asked how much of their process was automated, answers ranged from \inlinequote{None of it's automated} to \inlinequote{25 -- 35\%}.


Six subjects indicated dis-satisfaction with the current level of automation. They made statements like 
\inlinequote{that's a big point that could be improved}.
Two additional subjects indicated that the process was all or mostly manual, making statements like: \inlinequote{It's still pretty manual.}

Seven subjects described hindrances to more automation.
Two subjects mentioned that analysts had insufficient time to generate automation. 
Subjects made comments like \inlinequote{Mainly just lack of time development time. We've just been very busy.}
Two subjects described a lack of personnel. 
For example one subject responded: \inlinequote{Probably just personnel gaps.}
Two subjects mentioned that the team did not have the necessary knowledge to automate tasks that should be automated. 
When asked why more had not been automated one of these two subjects said: \inlinequote{because nobody understands [tool]}. 


Two subjects hesitated about further automation.
One team lead did not want the automation to become a crutch.
Another subject (in a leadership role) had a philosophical concern: 

\blockquote{Hunting...should start where automation stops.
That's...the whole premise of hunting...
the sophisticated actor already bypassed all that [automation] and now you have to apply manual techniques to...find them...
It's good to automate [repetitive] things, maybe like deploying kit.
But I think most of the analytical work should...rely on human factor.}
 

\mysubparagraph{(2) Ineffective Automation} 
\label{loops_rq1}
A second issue with automation is its effectiveness. 
\cref{fig:CoarseInducedDiagram} includes two analysis loops: manual and automated. 
Out of the subjects that had detected adversaries on hunt missions (7/11), none of their examples described activity found by the automated alert loop.
One team lead said: 
\inlinequote{
[The adversary] was almost exclusively found by analysts observing the data for living off the land techniques or new zero days.\footnote{``Living off the land'': using the victim's own tools instead of downloading external tools. ``Zero day'' vulnerabilities: unknown to network defenders.} I would say that we found at least three or four zero days, which, couldn't have been detected by [automation or intelligence reports].} 
A member of leadership said: \inlinequote{
[What] we find are mostly analytical kind of behavioral activity. Like an analyst spots.}


\myparagraph{Solutions:}
Five subjects suggested opportunities for greater automation.
Two subjects said baselining took too long and should be automated. 
One said: \inlinequote{Have the baseline step be an automatic step. ... where baseline is also is almost automatic. ... then you can kind of shift the window baseline out from two to three days...to a five minute thing.}
Two other subjects discussed equipment preparation tasks. One suggested:
\inlinequote{It would be cool to automate the indicator portion, other than us having to manually feed in that stuff.}
Subjects also suggested automating endpoint log collection and automating the reporting of important artifacts and their contexts to the team.

\subsubsection{Improving Data Collection}
\label{data_rq2}

\myparagraph{Problems:}
A successful threat hunt needs timely and sufficient data.
Subjects noted challenges in both dimensions.

\mysubparagraph{(1) Delayed data collection}
Many teams would only begin collecting data when the team arrived on site. 
This created an issue because the team was forced to start hunting before much baseline data had been collected. 
Without data, threat hunting is virtually impossible or as one experienced team lead said: \inlinequote{you can't find anything without data}.
Seven subjects either mention this issue
or said that they have learned to deploy sensors ahead of time to combat it.
One team lead said: \inlinequote{We did a threat hunting engagement...for
about two weeks, but we plugged sensors in on Day One of those two weeks.
When you do that, you don't have a baseline of what even one work week looks like much less a couple...we 
...
didn't have the 
baseline to actually do real analysis on the network.}

\mysubparagraph{(2) Un-prioritized Data Collection}
Some TH teams collected too much data or the wrong type of data. 
This extraneous data obscures potentially relevant data. 
One experienced member of leadership said: \inlinequote{I know with [former TH organization]...we
would ask for a ton of stuff. Give us this, and that...and
for no reason, right? ...
Just because you have all the data does not necessarily mean that you're going to be more effective doing your hunt. I think it's the opposite and that's kind of the mindset that needs to change.}

Another subject, a team lead, gave an example of a mission where so much data was collected that the servers containing the data stopped behaving properly.
The subject reported that in this environment: 
\inlinequote{if you tried to follow the checklist you ended up getting really confusing results because the logs that were coming in were non-deterministic ... 
trying to stick to [the checklist] was counterproductive because if you tried to stick to it, the data made less sense.} 

\myparagraph{Solutions:}
Five subjects had been on teams that had experienced the issue of lacking data when the team began analysis.
These subjects described a solution that worked for their teams: Weeks before the TH teams' arrival, a predeployment team would install sensors on the network. 
One subject said:
\inlinequote{I'll bring up a problem that we had. \textless laugh\textgreater our first one that we've worked through, which was a huge improvement, was the prep work in advance, right?
[Before]
we just set sensors the day that we arrived and that was terrible.}

The issue of unhelpful data being collected seemed to be difficult.
No subjects mentioned having solved this issue.
One experienced member of leadership suggested more precise scoping as a possible solution. 
They said: \inlinequote{We go into these environments, you see a lot of guest networks, or IOT devices, like cameras and security badges and stuff. It's stuff that we don't necessarily need to monitor
...and then it just fills up our sensors
...[queries] take forever to get results.
Let's concentrate on one thing, but do more in depth work.
Concentrate on quality, I guess, versus quantity.}

\subsubsection{Improving Process Documentation}
\label{detail_rq2}


Subjects expressed opposing views on their team's process documentation, as well as what would constitute good process documentation.
We therefore do not present this theme in terms of problem-solution, but rather in terms of these opposing views. 
Conflicts in these views are reminiscent of the "contradictions" observed in SOCs by Sundaramurthy \etal\cite{sundaramurthy_turning_2016}.
These views may be influenced by the level of detail in process documentation on each team. 
One organization had a detailed checklist. 
Another had a similarly detailed process but allowed team leads to adapt it per deployment.

\myparagraph{View--More detailed process documentation:}
Four subjects requested or supported more detailed process documentation.
\inlinequote{I think having the [detailed process documentation] will help [new members] be more effective, faster, because it will give them a guideline of what to do. So instead of sitting there not knowing what the first step to take, at least they know a general network to look in. So instead of sitting there: `I don't even know where to start.' It's: `I'm looking for connections on odd ports. That's my first thing. And that's how I'm gonna learn to build my first queries.'
}
More documentation and process rigidity was considered helpful for newer members.
One analyst described the creation of a spreadsheet documenting the task that had to be completed on a TH mission: \inlinequote{We didn't really know what we were looking for before spreadsheet. ... I took the idea of the spreadsheet by looking at Mitre \attack and seeing what we could look for. ... so it helped [new members] focus on each task at a time.}

\myparagraph{View--Less detailed process documentation:}
In contrast, four subjects spoke in favor of less detailed process documentation.
One subject said their team's process was too detailed: \inlinequote{For the current experience level of [my team], I would like it to see it a bit more vague... [subject gave an example of a VPN checklist item for a network without VPN]
...It's too specific, you know?
...[We should] make that more abstract and say like, `Hey, why don't [you] just look for any remote access software?'}.
One organization addresses this conflict by giving team leads autonomy:
\inlinequote{[Team leads] are responsible for their mission...They are given leeway to adjust as things happen.} 
When asked how often team leads adjust the process, this subject said: \inlinequote{I'd say 25 to 50\% of the time}.

In support of this view, one team lead gave an example of a mission where the process felt over-specified and switching to a more abstract process improved their performance. 
%
Another subject in leadership spoke to the importance of flexibility: \inlinequote{I dislike making rigid guidelines
...I think the goals of each engagement may be different based on the partner, the threat actor, the geopolitics 
...I think making it too strict makes it so that we lose some of the flexibility to do the missions that have the most impact or that we can't meet the goals we want based on a checklist that wasn't developed for them.}

\subsubsection{Turnover}
\label{turnover_rq2}
\vspace{-0.03cm}
\myparagraph{Problem:}
Three subjects mentioned the negative effects of turnover. 
One subject said: 
\inlinequote{our biggest problem right now is turnover ... [at] any given time, you maybe have a quarter to a third of your team, [that] has been on more than one mission}. 
They mentioned a case where their entire team was replaced with new members. 

\myparagraph{Solutions:}
Two possible solutions were offered by subjects: pairing members and improving process documentation.

\mysubparagraph{(1) Pairing Members}
Six subjects recommended pairing new members with expert members to improve the integration of newer members. 
One less experienced subject found it quite helpful, saying \inlinequote{[Observing more experienced members] was nice because I was basically able to look over the shoulder over all the missions that have happened so far and see where I fit in the puzzle piece of Threat Hunting.}.

Two more experienced subjects agreed that pairing members can help, but voiced warnings.
One spoke about short- vs. long-term payoffs:
\inlinequote{It's a balance between, do we need to succeed in this mission or do we need train our junior analysts to succeed in the next mission?}
The other emphasized that pairing requires learner engagement: \inlinequote{If
the new member isn't particularly motivated then they're just gonna be staring at a screen and not learning anything
...it would be nice to
...have
...a side saddle process...[where] let's both do it at the same time separately [and] you show me how to do it.}

\mysubparagraph{(2) Process Documentation}
\label{documentation_rq2}
Five subjects described good process documentation as important for assisting new members. 
Some subjects felt that process documentation helps a team meet a minimum standard, 
\eg
\inlinequote{I think defining...baselines...really well so you have the minimum standards and a clear task list...
What [are] the 10 steps that you absolutely will do before going deeper?
And if you have that well defined and built into your tools with automations when possible, that makes it...easier for people to come on board and get up to speed.} 
Another subject said: \inlinequote{Usually the less experienced ones are better at following the process, mostly because they don't know any better to do anything else.}

\section{Discussion \& Future Work} \label{dis}


Our work has two audiences: 
  operations teams
  and
  researchers.
We discuss TH among other cybersecurity operations (\cref{cyberops}), 
then share implications of our study for TH teams (\cref{team_recs_dis})
  and
  opportunities for future research (\cref{sec:Discussion-FutureWork}).

\subsection{Threat Hunt and Other CyberOps} \label{cyberops}
\new{As discussed in~\cref{background}, TH work is adjacent to Incident Response (IR) and Security Operation Centers (SOCs).
The experiences of our DHS TH team member subjects thus echo, yet are distinct, from the experiences of IR and SOC teams.}

\new{The DHS TH teams described many challenges shared by IR teams.
Our subjects said that measuring TH team effectiveness is difficult due to lack of feedback; IR teams have a similar issue~\cite{kleij_computer_2017}.
IR analysts express similar concerns about automation hindering analysis capability~\cite{Nyre-Yu_determining_2019}. 
Data access is an issue for both TH and IR teams\cite{grispos_security_2015}.
However, IR teams differ from our TH subjects in the data they have about the adversary. One reason why models such as \attack and Kill Chain are popular is that they show IR teams the ``next step'' when investigating an incident. For TH teams (and SOCs), there is no confirmed incident, so a TH team must investigate everything, hoping a clue was dropped along the kill chain.}

\new{With respect to SOCs, the studied TH teams operate as a third-party ``check'' on the SOCs' networks.
Our subjects thus undertake similar activities, but with different information.
The DHS TH teams get intelligence from other government entities, much of which cannot be shared with private sector SOCs.
The DHS TH teams also set up their own sensors to check for blind spots in the SOCs' surveillance.
Beyond these informational strengths lie weaknesses: DHS TH teams struggle to baseline because they are not on the network for long.
This may be why some subjects and other TH researchers that believe that TH can never be fully automated \cite{bianco_simple_nodate,miazi_design_2017} --- automation requires a baseline that third-party hunters lack.} 

\subsection{Recommendations for Threat Hunt Teams} \label{team_recs_dis}

In \cref{rq2} we presented challenges and potential solutions.
Now we synthesize our observations into three recommendations. 

\myparagraph{(1) Improve Planning.}
Although many subjects focused on the TH activities \textit{after} deploying to a customer site, about half of our unified TH process (\cref{fig:CoarseInducedDiagram}) occurs before deployment.
We were surprised by the range of planning activities and formality reported by subjects.
We recommend that all TH teams create mission planning documents \textbf{and} share the objectives with everyone on the team.
Following the advice of one subject: \inlinequote{plan the mission...your high-level goals...and push that into...specific tools and analytics.}

\myparagraph{(2) Revisit the Automated Alert Loop.} \label{sec:Discussion-Recs-AutomatedLoop} 
In~\cref{rq1_walkthrough} we described the two analysis loops applied by the TH teams we studied (see~\cref{fig:CoarseInducedDiagram}).
However, in~\cref{loops_rq1} our subjects reported that their automated alert loops rarely find adversary activity.
There are many possible causes, including
  redundant automation (\eg already applied by SOC teams),
  human error (\eg important alert is forgotten and not uploaded to tool)
  and
  absent automation (\eg zero-day exploits).
We recommend either making the automated alert loop more effective or reducing the resources devoted to it. 

The ineffectiveness of automated alerts might be a flaw in threat intelligence. 
One analyst, describing adversary detection on a previous mission, was surprised that there was no alert for the activity.
They said: \inlinequote{I don't think there was an alert on it. Which is odd thinking about it now...you definitely would think that [tool] would contain some...but I don't think there was.}
Eight subjects claimed that their intelligence component supplies the team with at least some of the alerting rules used during a hunt. 
Perhaps TH teams should revisit the trust they place in these turnkey automated rules. 

If TH teams do not revisit their automated alert strategy, then they might re-evaluate how many resources they invest in it. 
For example, one subject estimated that intelligence/alerting only accounts for \inlinequote{maybe a tenth} of adversary detections.
If the amount of resources put into this analytic loop exceeds 10\%, this could indicate an inefficiency.


\myparagraph{(3) Formalize apprenticeship.} \label{sec:Discussion-Recs-Apprenticeship} 
Many subjects voiced concern about the quality of existing TH training and certifications~\cref{expertise_rq2}.
Some teams addressed this by pairing new and experienced members (\cref{turnover_rq2}). 
These teams appear to be reinventing the concept of \textit{apprenticeship}, which is an educational strategy for disciplines that are more art than science~\cite{noauthor_cambridge_2014}.
There is much literature on the virtues and shortcomings of apprenticeship~\cite{borg_apprenticeship_2004,ryan_is_1998} and apprenticeship is currently being used with success in cybersecurity~\cite{smith_computing_2020,stoker_building_2021}.
Since subjects seem to agree that pairing is a good way to integrate new members, teams may benefit from a more robust apprenticeship program.
According to subjects, this should include providing time while on mission, since hunting with trainees takes longer.
We recommend ensuring trainees not just watch others hunt, but rather interact with data and tools under supervision (the ``side saddle'' process suggested by one subject). 

\subsection{Future Work} \label{sec:Discussion-FutureWork}

Threat Hunt is a young cybersecurity discipline.
Our exploratory study suggests many beneficial directions of study.


\myparagraph{Tailor automation to needs:}
We reported subjects' suggestions for automation in~\cref{automation_rq2}. 
The most popular suggestions were automating baselining and recurring equipment set-up tasks. 
In contrast, most of the automation in the academic literature focuses on automating the entire TH process~\cite{wei_deephunter_2021,karuna_automating_2021,horta_neto_cyber_2020,milajerdi_poirot_2019}.
This class of automation may address the current shortcomings of the automated analysis loop. 
However, it does not address the needs verbalized by our subjects.  

\myparagraph{Does automation hinder analysts?}
In \cref{automation_rq2}, one team lead and one subject in a leadership role expressed concerns that automation would lead analysts to not think for themselves. 
Measuring whether, and under what conditions, automation reduces team effectiveness would provide useful data for TH decision-makers.



\myparagraph{Process formalism or flexibility?}
In \cref{detail_rq2}, subjects request more detailed TH process documentation.
However, in that section, an example is given of a process that was too precise, becoming a hindrance to the team. 
A second subject indicates that they wish the process were more vague. 
Some teams deal with this tension by allowing team leads to make process changes.
Not all organizations require their team leads to have TH experience so not all team leads will be capable of making these decisions. 
It is possible that this tension is inherent to a creative and open-ended activity, but tracking and measuring it may be helpful. 



\myparagraph{Process evaluation:}
Subjects did not agree on metrics for measuring TH team effectiveness. 
The goal of TH is to reduce adversary dwell time~\cite{van_os_tahiti_2018}.
However, this metric may not be applicable to the TH teams we studied because they act in a third-party capacity and do not necessarily revisit organizations.
Additionally, organizations may be compromised so infrequently that measuring dwell times does not provide a TH team with actionable feedback. 

Other metric recommendations include
  security improvement~\cite{ewing_endgame_nodate},
  and
  risk reduction~\cite{gutzwiller_gaps_2020}.
These metrics measure whether a TH team added value by identifying vulnerabilities or security blind spots. 
An objective measure of this kind of effectiveness 
may help evaluate the quality of a TH process.
A final proposal is Bianco's Hunt Maturity Model~\cite{bianco_simple_nodate}, which measures a team's capabilities rather than its products.
None of these metrics has been systematically evaluated in practice.

\section{Conclusion}
\label{sec:concl}

Many organizations have recently adopted Threat Hunting as a way to detect adversaries who have infiltrated their networks undetected.
There is little academic literature on Threat Hunting processes, and no description of the Threat Hunting process as performed by US government hunt teams working on third-party networks.
In this work, we provide the first description of the US government Threat Hunting process model as practiced by teams in the US Department of Homeland Security.
We found that these teams have different processes than those reported in prior literature, differing from both private-sector internal hunt teams and other government teams.
We provide a novel model of the Threat Hunt process, complemented with a set of open problems and possible solutions suggested by Threat Hunt practitioners.
Much work remains: in the short term, these process recommendations can be implemented and tested; in the long term, experiments in this sensitive context remain an open challenge. 

\ifANONYMOUS
\else
\section{Acknowledgments}
We acknowledge support from the Purdue Military Research Institute and the US Coast Guard.
We thank G. Cramer for assistance in coding, and A. Kazerouni for feedback on the manuscript.
We thank the CCS and USENIX reviewers for constructive criticism.
\fi

\raggedbottom
\pagebreak

\bibliographystyle{plainurl}

\bibliography{paper}

\begin{thebibliography}{100}

\bibitem{noauthor_cambridge_2014}
{\em The {Cambridge} {Handbook} of the {Learning} {Sciences}}.
\newblock Cambridge {Handbooks} in {Psychology}. Cambridge University Press, 2
  edition, 2014.
\newblock \href {https://doi.org/10.1017/CBO9781139519526}
  {\path{doi:10.1017/CBO9781139519526}}.

\bibitem{noauthor_kill_2014}
A "{Kill} {Chain}" analysis of the 2013 target data breach, March 2014.
\newblock URL:
  \url{https://www.commerce.senate.gov/services/files/24d3c229-4f2f-405d-b8db-a3a67f183883}.

\bibitem{noauthor_gaining_2015}
Gaining {The} {Advantage}, 2015.
\newblock URL:
  \url{https://www.lockheedmartin.com/content/dam/lockheed-martin/rms/documents/cyber/Gaining_the_Advantage_Cyber_Kill_Chain.pdf}.

\bibitem{noauthor_lessons_2015}
Lessons to {Learn} from the {OPM} {Breach}, June 2015.
\newblock URL:
  \url{https://www.tenable.com/blog/lessons-to-learn-from-the-opm-breach}.

\bibitem{noauthor_u_2015}
({U}) {Combat} {Mission} {Teams} and {Cyber} {Protection} {Teams} {Lacked}
  {Adequate} {Capabilities} and {Facilities} to {Perform} {Missions}.
\newblock Technical Report DODIG-2016-026, Inspector general of the US
  Department of Defense, November 2015.
\newblock URL:
  \url{https://media.defense.gov/2017/Sep/15/2001810710/-1/-1/1/DODIG-2016-026%20(REDACTED).PDF}.

\bibitem{noauthor_lessons_2016}
Lessons from the {OPM} breach, September 2016.
\newblock URL:
  \url{https://gcn.com/cybersecurity/2016/09/lessons-from-the-opm-breach/316728/}.

\bibitem{noauthor_cybersecurity_2017}
Cybersecurity {Experts} {Hunting} for {Hackers}, April 2017.
\newblock URL:
  \url{https://www.nationaldefensemagazine.org/articles/2017/4/3/cybersecurity-experts-hunting-for-hackers}.

\bibitem{noauthor_national_2018}
National {Cyber} {Strategy}, September 2018.
\newblock URL:
  \url{https://trumpwhitehouse.archives.gov/wp-content/uploads/2018/09/National-Cyber-Strategy.pdf}.

\bibitem{noauthor_cost_2019}
The {Cost} of {Cybercrime}.
\newblock Technical report, Ponemon Institute LLC and Accenture, 2019.

\bibitem{noauthor_cost_2020}
Cost of a {Data} {Breach} {Report} 2020.
\newblock Technical report, IBM Security, July 2020.
\newblock URL:
  \url{https://www.ibm.com/security/digital-assets/cost-data-breach-report/1Cost%20of%20a%20Data%20Breach%20Report%202020.pdf}.

\bibitem{noauthor_guide_2020}
{Guide} {To} {Cyber} {Threat} {Hunting}, 2020.
\newblock URL:
  \url{https://www.tylertech.com/services/ndiscovery/nDiscovery-Threat-Hunting.pdf}.

\bibitem{noauthor_how_2020}
How {High} {Employee} {Turnover} {Poses} {Increased} {Cyber} {Security} {Risk},
  November 2020.
\newblock URL: \url{https://copperbandtech.com/high-employee-turnover/}.

\bibitem{noauthor_6_2021}
6 {U}.{S}.{C}. §659, December 2021.

\bibitem{noauthor_attack_2021}
Attack vs. {Data}: {What} {You} {Need} to {Know} {About} {Threat} {Hunting}
  {\textbar} {Rapid7} {Blog}, March 2021.
\newblock URL:
  \url{https://www.rapid7.com/blog/post/2021/03/25/attack-vs-data-what-you-need-to-know-about-threat-hunting/}.

\bibitem{noauthor_cyber_2021}
Cyber {Security} {Analyst} {Demographics} {And} {Statistics} {In} {The} {US},
  December 2021.
\newblock URL:
  \url{https://www.zippia.com/cyber-security-analyst-jobs/demographics/}.

\bibitem{noauthor_executive_2021}
Executive {Order} 14028: {Improving} the {Nation}’s {Cybersecurity}, May
  2021.
\newblock URL:
  \url{https://www.whitehouse.gov/briefing-room/presidential-actions/2021/05/12/executive-order-on-improving-the-nations-cybersecurity/}.

\bibitem{noauthor_cisa_2022}
{CISA} {Strategic} {Plan} 2023-2025.
\newblock page~37, September 2022.

\bibitem{noauthor_cyber_2022}
{CYBER} 101: {Hunt} {Forward} {Operations}, November 2022.
\newblock URL:
  \url{https://www.cybercom.mil/Media/News/Article/3218642/cyber-101-hunt-forward-operations/https%3A%2F%2Fwww.cybercom.mil%2FMedia%2FNews%2FArticle%2F3218642%2Fcyber-101-hunt-forward-operations%2F}.

\bibitem{noauthor_falcon_2022}
Falcon {Overwatch}: {Managed} \& {Proactive} {Threat} {Hunting} {\textbar}
  {CrowdStrike}, 2022.
\newblock URL:
  \url{https://www.crowdstrike.com/endpoint-security-products/falcon-overwatch-threat-hunting/}.

\bibitem{cybersecurity_and_infastructure_security_agency_malicious_2022}
Malicious {Cyber} {Actors} {Continue} to {Exploit} {Log4Shell} in {VMware}
  {Horizon} {Systems}.
\newblock Cybersecurity {Advisory} AA22-174A, Cybersecurity and Infastructure
  Security Agency, June 2022.

\bibitem{noauthor_talos_2022}
Talos {Incident} {Response}, 2022.
\newblock URL: \url{https://talosintelligence.com/incident_response/hunting}.

\bibitem{noauthor_threat_2022}
Threat {Hunting}, 2022.
\newblock URL:
  \url{https://www.boozallen.com/expertise/cybersecurity/threat-hunting.html}.

\bibitem{noauthor_united_2022}
The {United} {States} {Government} {Manual}, 2022.
\newblock URL:
  \url{https://www.govinfo.gov/content/pkg/GOVMAN-2022-12-31/xml/GOVMAN-2022-12-31.xml}.

\bibitem{singleton_x-force_2022}
X-{Force} {Threat} {Intelligence} {Index} 2022.
\newblock Technical report, IBM Security, February 2022.
\newblock URL: \url{https://www.ibm.com/downloads/cas/ADLMYLAZ}.

\bibitem{noauthor_inquest_2023}
{InQuest} {Labs} - {IOCDB} - {InQuest}.net, 2023.
\newblock URL: \url{https://labs.inquest.net}.

\bibitem{noauthor_national_2023}
National {Cyber} {Strategy}, March 2023.
\newblock URL:
  \url{https://www.whitehouse.gov/wp-content/uploads/2023/03/National-Cybersecurity-Strategy-2023.pdf}.

\bibitem{agarwal_cyber_2021}
Anchit Agarwal, Himdweep Walia, and Himanshu Gupta.
\newblock Cyber {Security} {Model} for {Threat} {Hunting}.
\newblock In {\em 2021 9th {International} {Conference} on {Reliability},
  {Infocom} {Technologies} and {Optimization} ({Trends} and {Future}
  {Directions}) ({ICRITO})}, pages 1--8, September 2021.
\newblock \href {https://doi.org/10.1109/ICRITO51393.2021.9596199}
  {\path{doi:10.1109/ICRITO51393.2021.9596199}}.

\bibitem{alharbi_security_nodate}
Norah Alharbi.
\newblock {\em A {Security} {Operation} {Center} {Maturity} {Model}
  ({SOC}-{MM}) in the {Context} of {Newly} {Emerging} {Cyber} {Threats}}.
\newblock Ph.{D}., The Claremont Graduate University, United States --
  California.
\newblock ISBN: 9798672151229.

\bibitem{alkhadra_solar_2021}
Rahaf Alkhadra, Joud Abuzaid, Mariam AlShammari, and Nazeeruddin Mohammad.
\newblock Solar winds hack: {In}-depth analysis and countermeasures.
\newblock In {\em 2021 12th {International} {Conference} on {Computing}
  {Communication} and {Networking} {Technologies} ({ICCCNT})}, pages 1--7.
  IEEE, 2021.

\bibitem{pennington_getting_2019}
{Applebaum, Andy}, {Nickels, Katie}, {Schulz, Tim}, {Strom, Blake}, and
  {Wunder, John}.
\newblock Getting {Started} with {ATT}\&{CK}, October 2019.
\newblock URL:
  \url{https://www.mitre.org/sites/default/files/2021-11/getting-started-with-attack-october-2019.pdf}.

\bibitem{araujo_evidential_2021}
Frederico Araujo, Dhilung Kirat, Xiaokui Shu, Teryl Taylor, and Jiyong Jang.
\newblock Evidential {Cyber} {Threat} {Hunting}.
\newblock {\em arXiv:2104.10319 [cs]}, April 2021.
\newblock arXiv: 2104.10319.

\bibitem{bartlett_organizational_nodate}
James~E Bartlett, Joe~W Kotrlik, and Chadwick~C Higgins.
\newblock Organizational {Research}: {Determining} {Appropriate} {Sample}
  {Size} in {Survey} {Research}.

\bibitem{bianco_simple_nodate}
David~J Bianco.
\newblock A {Simple} {Hunting} {Maturity} {Model}.
\newblock URL:
  \url{http://detect-respond.blogspot.com/2015/10/a-simple-hunting-maturity-model.html}.

\bibitem{davidjbianco_enterprise_2013}
David~J Bianco.
\newblock Enterprise {Detection} \& {Response}: {The} {Pyramid} of {Pain},
  March 2013.
\newblock URL:
  \url{https://detect-respond.blogspot.com/2013/03/the-pyramid-of-pain.html}.

\bibitem{sqrrl_team_framework_nodate}
David~J Bianco.
\newblock A {Framework} for {Cyber} {Threat} {Hunting} {Part} 1: {The}
  {Pyramid} of {Pain}, 2015.
\newblock URL:
  \url{https://www.threathunting.net/files/A%20Framework%20for%20Cyber%20Threat%20Hunting%20Part%201_%20The%20Pyramid%20of%20Pain%20_%20Sqrrl.pdf}.

\bibitem{birt_member_2016}
Linda Birt, Suzanne Scott, Debbie Cavers, Christine Campbell, and Fiona Walter.
\newblock Member {Checking}: {A} {Tool} to {Enhance} {Trustworthiness} or
  {Merely} a {Nod} to {Validation}?
\newblock {\em Qualitative Health Research}, 26(13):1802--1811, November 2016.
\newblock \href {https://doi.org/10.1177/1049732316654870}
  {\path{doi:10.1177/1049732316654870}}.

\bibitem{bledsoe_how_2023}
Everett Bledsoe.
\newblock How {Often} {Do} {Military} {Families} {Move}? {Why} {They} {Move}
  {So} {Much}?, January 2023.
\newblock URL:
  \url{https://www.thesoldiersproject.org/how-often-do-military-families-move/}.

\bibitem{borg_apprenticeship_2004}
Michaela Borg.
\newblock The apprenticeship of observation.
\newblock {\em Elt Journal}, 58:274--276, July 2004.
\newblock \href {https://doi.org/10.1093/elt/58.3.274}
  {\path{doi:10.1093/elt/58.3.274}}.

\bibitem{brink_quantifying_2017}
Derek~E Brink.
\newblock Quantifying the {Value} of {Time} in {Cyber}-{Threat} {Detection} and
  {Response}.
\newblock Technical Report 15218, Aberdeen Group, January 2017.

\bibitem{bynum_cyber_2019}
Jon~R Bynum.
\newblock Cyber {Threat} {Hunting}.
\newblock Master's thesis, Utica College, 2019.

\bibitem{chenail_interviewing_2011}
Ronald~J Chenail.
\newblock Interviewing the {Investigator}: {Strategies} for {Addressing}
  {Instrumentation} and {Researcher} {Bias} {Concerns} in {Qualitative}
  {Research}.
\newblock {\em The Qualitative Report}, 16(1):8, January 2011.

\bibitem{cho_capturing_2020}
Selina~Y. Cho, Jassim Happa, and Sadie Creese.
\newblock Capturing {Tacit} {Knowledge} in {Security} {Operation} {Centers}.
\newblock {\em IEEE Access}, 8:42021--42041, 2020.
\newblock \href {https://doi.org/10.1109/ACCESS.2020.2976076}
  {\path{doi:10.1109/ACCESS.2020.2976076}}.

\bibitem{cichonski_computer_2012}
Paul Cichonski, Tom Millar, Tim Grance, and Karen Scarfone.
\newblock Computer {Security} {Incident} {Handling} {Guide} : {Recommendations}
  of the {National} {Institute} of {Standards} and {Technology}.
\newblock Technical Report NIST SP 800-61r2, National Institute of Standards
  and Technology, August 2012.
\newblock \href {https://doi.org/10.6028/NIST.SP.800-61r2}
  {\path{doi:10.6028/NIST.SP.800-61r2}}.

\bibitem{de_gramatica_it_2015}
Martina De~Gramatica, Fabio Massacci, Woohyun Shim, Alessandra Tedeschi, and
  Julian Williams.
\newblock {IT} {Interdependence} and the {Economic} {Fairness} of
  {Cybersecurity} {Regulations} for {Civil} {Aviation}.
\newblock {\em IEEE Security \& Privacy}, 13(5):52--61, September 2015.
\newblock Conference Name: IEEE Security \& Privacy.
\newblock \href {https://doi.org/10.1109/MSP.2015.98}
  {\path{doi:10.1109/MSP.2015.98}}.

\bibitem{dodgson2019reflexivity}
Joan~E Dodgson.
\newblock Reflexivity in qualitative research.
\newblock {\em Journal of Human Lactation}, 35(2):220--222, 2019.

\bibitem{donalds_cybersecurity_2020}
Charlette Donalds and Kweku-Muata Osei-Bryson.
\newblock Cybersecurity compliance behavior: {Exploring} the influences of
  individual decision style and other antecedents.
\newblock {\em International Journal of Information Management}, 51:102056,
  April 2020.
\newblock \href {https://doi.org/10.1016/j.ijinfomgt.2019.102056}
  {\path{doi:10.1016/j.ijinfomgt.2019.102056}}.

\bibitem{ec-council_what_2022}
EC-Council.
\newblock What {Is} the {Pyramid} of {Pain}, and {Why} {Is} {It} {Important} in
  {Threat} {Detection}?, October 2022.
\newblock URL:
  \url{https://www.eccouncil.org/cybersecurity-exchange/threat-intelligence/pyramid-pain-threat-detection/}.

\bibitem{everson_network_2020}
Douglas Everson and Long Cheng.
\newblock Network {Attack} {Surface} {Simplification} for {Red} and {Blue}
  {Teams}.
\newblock In {\em 2020 {IEEE} {Secure} {Development} ({SecDev})}, pages 74--80,
  September 2020.
\newblock \href {https://doi.org/10.1109/SecDev45635.2020.00027}
  {\path{doi:10.1109/SecDev45635.2020.00027}}.

\bibitem{ewing_endgame_nodate}
Paul Ewing and Devon Kerr.
\newblock {\em The {Endgame} {Guide} to {Threat} {Hunting}}.
\newblock URL:
  \url{https://cyber-edge.com/resources/the-endgame-guide-to-threat-hunting/}.

\bibitem{fulton_vulnerability_2023}
Kelsey~R Fulton, Samantha Katcher, Kevin Song, Marshini Chetty, Michelle~L
  Mazurek, Chlo{\'e} Messdaghi, and Daniel Votipka.
\newblock Vulnerability discovery for all: Experiences of marginalization in
  vulnerability discovery.
\newblock In {\em 32nd USENIX Security Symposium (USENIX Security). USENIX
  Association}, 2023.

\bibitem{gao_enabling_2021}
Peng Gao, Fei Shao, Xiaoyuan Liu, Xusheng Xiao, Zheng Qin, Fengyuan Xu, Prateek
  Mittal, Sanjeev~R. Kulkarni, and Dawn Song.
\newblock Enabling {Efficient} {Cyber} {Threat} {Hunting} {With} {Cyber}
  {Threat} {Intelligence}.
\newblock In {\em 2021 {IEEE} 37th {International} {Conference} on {Data}
  {Engineering} ({ICDE})}, pages 193--204, April 2021.
\newblock ISSN: 2375-026X.
\newblock \href {https://doi.org/10.1109/ICDE51399.2021.00024}
  {\path{doi:10.1109/ICDE51399.2021.00024}}.

\bibitem{grispos_security_2015}
George Grispos, William~Bradley Glisson, and Tim Storer.
\newblock Security incident response criteria: {A} practitioner's perspective.
\newblock {\em CoRR}, abs/1508.02526, 2015.
\newblock URL: \url{http://arxiv.org/abs/1508.02526}, \href
  {http://arxiv.org/abs/1508.02526} {\path{arXiv:1508.02526}}.

\bibitem{guest_how_2006}
Greg Guest, Arwen Bunce, and Laura Johnson.
\newblock How {Many} {Interviews} {Are} {Enough}?: {An} {Experiment} with
  {Data} {Saturation} and {Variability}.
\newblock {\em Field Methods}, 18(1):59--82, February 2006.
\newblock \href {https://doi.org/10.1177/1525822X05279903}
  {\path{doi:10.1177/1525822X05279903}}.

\bibitem{guest_applied_2012}
Greg Guest, Kathleen MacQueen, and Emily Namey.
\newblock {\em Applied {Thematic} {Analysis}}.
\newblock SAGE Publications, Inc., 2455 Teller Road, Thousand
  Oaks California 91320 United States, 2012.
\newblock \href {https://doi.org/10.4135/9781483384436}
  {\path{doi:10.4135/9781483384436}}.

\bibitem{gutzwiller_gaps_2020}
Robert Gutzwiller, Josiah Dykstra, and Bryan Payne.
\newblock Gaps and {Opportunities} in {Situational} {Awareness} for
  {Cybersecurity}.
\newblock {\em Digital Threats: Research and Practice}, 1(3):1--6, September
  2020.
\newblock \href {https://doi.org/10.1145/3384471} {\path{doi:10.1145/3384471}}.

\bibitem{horta_neto_cyber_2020}
Antonio~José Horta~Neto and Anderson Fernandes Pereira~dos Santos.
\newblock Cyber {Threat} {Hunting} {Through} {Automated} {Hypothesis} and
  {Multi}-{Criteria} {Decision} {Making}.
\newblock In {\em 2020 {IEEE} {International} {Conference} on {Big} {Data}
  ({Big} {Data})}, pages 1823--1830, December 2020.
\newblock \href {https://doi.org/10.1109/BigData50022.2020.9378213}
  {\path{doi:10.1109/BigData50022.2020.9378213}}.

\bibitem{huck_wake_2022}
Jan Huck and Frank Breitinger.
\newblock Wake {Up} {Digital} {Forensics}’ {Community} and {Help} {Combat}
  {Ransomware}.
\newblock {\em IEEE Security \& Privacy}, 20(4):61--70, July 2022.
\newblock Conference Name: IEEE Security \& Privacy.
\newblock \href {https://doi.org/10.1109/MSEC.2021.3137018}
  {\path{doi:10.1109/MSEC.2021.3137018}}.

\bibitem{jacobs_classification_2013}
Pierre Jacobs, Alapan Arnab, and Barry Irwin.
\newblock Classification of {Security} {Operation} {Centers}.
\newblock In {\em 2013 {Information} {Security} for {South} {Africa}}, pages
  1--7, Johannesburg, South Africa, August 2013. IEEE.
\newblock \href {https://doi.org/10.1109/ISSA.2013.6641054}
  {\path{doi:10.1109/ISSA.2013.6641054}}.

\bibitem{jason_chaffetz_opm_nodate}
{Jason Chaffetz}, {Mark Meadows}, and {Will Hurd}.
\newblock The {OPM} {Data} {Breach}: {How} the {Government} {Jeopardized} {Our}
  {National} {Security} for {More} than a {Generation}.
\newblock Majority {Staff} {Report}, U.S. House of Representatives Committee on
  Oversight and Government Reform.

\bibitem{karuna_automating_2021}
Prakruthi Karuna, Erik Hemberg, Una-May O'Reilly, and Nick Rutar.
\newblock Automating {Cyber} {Threat} {Hunting} {Using} {NLP}, {Automated}
  {Query} {Generation}, and {Genetic} {Perturbation}.
\newblock {\em arXiv:2104.11576 [cs]}, April 2021.
\newblock arXiv: 2104.11576.

\bibitem{kerner_lessons_2017}
Sean~Michael Kerner.
\newblock Lessons {Learned} from the {OPM} {Breach} {eSecurity} {Planet},
  December 2017.
\newblock URL:
  \url{https://www.esecurityplanet.com/threats/lessons-learned-from-the-opm-breach/}.

\bibitem{klaas-jan_stol_grounded_2015}
{Klaas-Jan Stol}, {Paul Ralph}, and {Brian Fitzgerald}.
\newblock Grounded {Theory} in {Software} {Engineering} {Research}: {A}
  {Critical} {Review} and {Guidelines}.
\newblock In {\em 2015 {IEEE}/{ACM} 37th {IEEE} {International} {Conference} on
  {Software} {Engineering}}, Florence, Italy, 2015. IEEE.

\bibitem{lee_hunter_2017}
Written~Rob Lee and Robert~M Lee.
\newblock The {Hunter} {Strikes} {Back}: {The} {SANS} 2017 {Threat} {Hunting}
  {Survey}, 2017.
\newblock URL:
  \url{https://www.sans.org/webcasts/threat-hunting-modernizing-detection-operations-2017-threat-hunting-survey-results-1-103767/}.

\bibitem{lee_who_2016}
Written Robert~M Lee and Rob Lee.
\newblock The {Who}, {What}, {Where}, {When}, {Why} and {How} of {Effective}
  {Threat} {Hunting}, February 2016.
\newblock URL:
  \url{https://www.sans.org/white-papers/who-what-where-when-why-how-effective-threat-hunting/}.

\bibitem{lemos_5_2016}
Robert Lemos.
\newblock 5 {Lessons} {Learned} {From} {OPM} {Data} {Breach}, September 2016.
\newblock URL:
  \url{https://www.eweek.com/security/5-revelations-from-opm-data-breach-report/}.

\bibitem{majid_success_2019}
M.~Majid and K~Ariffi.
\newblock Success {Factors} for {Cyber} {Security} {Operation} {Center} ({SOC})
  {Establishment}.
\newblock In {\em Proceedings of the {Proceedings} of the 1st {International}
  {Conference} on {Informatics}, {Engineering}, {Science} and {Technology},
  {INCITEST} 2019, 18 {July} 2019, {Bandung}, {Indonesia}}, Bandung, Indonesia,
  2019. EAI.
\newblock \href {https://doi.org/10.4108/eai.18-7-2019.2287841}
  {\path{doi:10.4108/eai.18-7-2019.2287841}}.

\bibitem{matrosov_stuxnet_2010}
Aleksandr Matrosov, Eugene Rodionov, David Harley, and Juraj Malcho.
\newblock Stuxnet under the microscope.
\newblock {\em ESET LLC (September 2010)}, 6, 2010.

\bibitem{mcclanahan_169th_2019}
Sarah~M. McClanahan.
\newblock 169th {Cyber} {Protection} {Team} – {Highly} {Capable}, {Always}
  {Ready}, November 2019.
\newblock URL:
  \url{https://news.maryland.gov/ng/2019/11/06/169th-cyber-protection-team-highly-capable-always-ready/}.

\bibitem{mciver_closing_2022}
Kathryn McIver.
\newblock Closing the {Revolving} {Door}.
\newblock Technical report, August 2022.
\newblock URL:
  \url{https://icitech.org/wp-content/uploads/2022/08/Closing-the-Revolving-Door.pdf}.

\bibitem{miazi_design_2017}
Md~Nazmus~Sakib Miazi, Mir Mehedi~A. Pritom, Mohamed Shehab, Bill Chu, and
  Jinpeng Wei.
\newblock The {Design} of {Cyber} {Threat} {Hunting} {Games}: {A} {Case}
  {Study}.
\newblock In {\em 2017 26th {International} {Conference} on {Computer}
  {Communication} and {Networks} ({ICCCN})}, pages 1--6, July 2017.
\newblock \href {https://doi.org/10.1109/ICCCN.2017.8038527}
  {\path{doi:10.1109/ICCCN.2017.8038527}}.

\bibitem{milajerdi_poirot_2019}
Sadegh~M. Milajerdi, Birhanu Eshete, Rigel Gjomemo, and V.~N. Venkatakrishnan.
\newblock {POIROT}: {Aligning} {Attack} {Behavior} with {Kernel} {Audit}
  {Records} for {Cyber} {Threat} {Hunting}.
\newblock {\em Proceedings of the 2019 ACM SIGSAC Conference on Computer and
  Communications Security}, pages 1795--1812, November 2019.
\newblock arXiv: 1910.00056.
\newblock \href {https://doi.org/10.1145/3319535.3363217}
  {\path{doi:10.1145/3319535.3363217}}.

\bibitem{milea_hypothesis_2017}
Demetrio Milea.
\newblock Hypothesis in {Threat} {Hunting}, July 2017.
\newblock URL:
  \url{https://medium.com/@demetriom/hypothesis-in-threat-hunting-4bea5446e34c}.

\bibitem{naseer_real-time_2021}
Ayesha Naseer, Humza Naseer, Atif Ahmad, Sean~B. Maynard, and Adil
  Masood~Siddiqui.
\newblock Real-time analytics, incident response process agility and enterprise
  cybersecurity performance: {A} contingent resource-based analysis.
\newblock {\em International Journal of Information Management}, 59:102334,
  August 2021.
\newblock \href {https://doi.org/10.1016/j.ijinfomgt.2021.102334}
  {\path{doi:10.1016/j.ijinfomgt.2021.102334}}.

\bibitem{military_family_advisory_network_effects_nodate}
Military Family~Advisory Network.
\newblock Effects of {Moving} on {Military} {Families}.
\newblock URL:
  \url{https://www.mfan.org/topic/moving-permanent-change-of-station/effects-of-moving-on-military-families/}.

\bibitem{nosco_industrial_2020}
Tim Nosco, Jared Ziegler, and Zechariah Clark.
\newblock The {Industrial} {Age} of {Hacking}.
\newblock In {\em Proceedings of the 29th {USENIX} {Security} {Symposium}},
  August 2020.

\bibitem{nyre-yu_observing_2019}
Megan Nyre-Yu, Robert~S. Gutzwiller, and Barrett~S. Caldwell.
\newblock Observing {Cyber} {Security} {Incident} {Response}: {Qualitative}
  {Themes} {From} {Field} {Research}.
\newblock {\em Proceedings of the Human Factors and Ergonomics Society Annual
  Meeting}, 63(1):437--441, November 2019.
\newblock \href {https://doi.org/10.1177/1071181319631016}
  {\path{doi:10.1177/1071181319631016}}.

\bibitem{Nyre-Yu_determining_2019}
Megan~M Nyre-Yu.
\newblock {Determining System Requirements for Human-Machine Integration in
  Cyber Security Incident Response}.
\newblock 10 2019.
\newblock \href {https://doi.org/10.25394/PGS.10014803.v1}
  {\path{doi:10.25394/PGS.10014803.v1}}.

\bibitem{oltsik_life_2019}
Jon Oltsik.
\newblock The {Life} and {Times} of {Cybersecurity} {Professionals} 2018.
\newblock {\em Research Report}, 2019.
\newblock URL:
  \url{https://www.esg-global.com/hubfs/pdf/ESG-ISSA-Research-Report-Life-of-Cybersecurity-Professionals-Apr-2019.pdf}.

\bibitem{pettigrew_making_2012}
James Pettigrew and Julie Ryan.
\newblock Making {Successful} {Security} {Decisions}: {A} {Qualitative}
  {Evaluation}.
\newblock {\em IEEE Security \& Privacy}, 10(1):60--68, January 2012.
\newblock Conference Name: IEEE Security \& Privacy.
\newblock \href {https://doi.org/10.1109/MSP.2011.128}
  {\path{doi:10.1109/MSP.2011.128}}.

\bibitem{rasheed_threat_2017}
Hussein Rasheed, Ali Hadi, and Mariam Khader.
\newblock Threat {Hunting} {Using} {GRR} {Rapid} {Response}.
\newblock In {\em 2017 {International} {Conference} on {New} {Trends} in
  {Computing} {Sciences} ({ICTCS})}, pages 155--160, October 2017.
\newblock \href {https://doi.org/10.1109/ICTCS.2017.22}
  {\path{doi:10.1109/ICTCS.2017.22}}.

\bibitem{ryan_is_1998}
Paul Ryan.
\newblock Is apprenticeship better? a review of the economic evidence.
\newblock {\em Journal of Vocational Education \& Training}, 50(2):289--325,
  June 1998.
\newblock \href {https://doi.org/10.1080/13636829800200050}
  {\path{doi:10.1080/13636829800200050}}.

\bibitem{sacher-boldewin_intelligent_2022}
Desiree Sacher-Boldewin and Eireann Leverett.
\newblock The {Intelligent} {Process} {Lifecycle} of {Active} {Cyber}
  {Defenders}.
\newblock {\em Digital Threats: Research and Practice}, 3(3):1--17, September
  2022.
\newblock \href {https://doi.org/10.1145/3499427} {\path{doi:10.1145/3499427}}.

\bibitem{sahin2023investigating}
Sena Sahin, Suood Al~Roomi, Tara Poteat, and Frank Li.
\newblock Investigating the password policy practices of website
  administrators.
\newblock In {\em 2023 IEEE Symposium on Security and Privacy (SP)}, pages
  552--569. IEEE, 2023.

\bibitem{saldana_coding_2021}
Johnny Saldaña.
\newblock The coding manual for qualitative researchers.
\newblock {\em The coding manual for qualitative researchers}, pages 1--440,
  2021.
\newblock Publisher: SAGE publications Ltd.

\bibitem{scarfone_hunters_nodate}
Karen Scarfone.
\newblock {\em The {Hunters} {Handbook}}.
\newblock ENDGAME \& Accenture.
\newblock URL:
  \url{https://cyber-edge.com/resources/the-hunters-handbook-endgames-guide-to-adversary-hunting/download/}.

\bibitem{schmid_military_2022}
Eric Schmid.
\newblock Military families often have to move every few years. {Critics} say
  it's disruptive and unnecessary, April 2022.
\newblock Section: Military.
\newblock URL:
  \url{https://wusfnews.wusf.usf.edu/military/2022-04-03/military-families-often-have-to-move-every-few-years-critics-say-its-disruptive-and-unnecessary}.

\bibitem{security_cyber_2022}
Cyborg Security.
\newblock Cyber {Threat} {Hunting} - {What} {Is} {It}, {Really}?, May 2022.
\newblock URL:
  \url{https://www.cyborgsecurity.com/blog/cyber-threat-hunting-what-is-it-really/}.

\bibitem{security_threat_2022}
Cyborg Security.
\newblock The {Threat} {Hunter}'s {Hypothesis}, March 2022.
\newblock URL:
  \url{https://www.cyborgsecurity.com/library/guides/the-threat-hunters-hypothesis-2/}.

\bibitem{smith_computing_2020}
Sally Smith, Ella Taylor-Smith, Khristin Fabian, Matthew Barr, Tessa Berg,
  David Cutting, James Paterson, Tiffany Young, and Mark Zarb.
\newblock Computing degree apprenticeships: {An} opportunity to address gender
  imbalance in the {IT} sector?
\newblock In {\em 2020 {IEEE} {Frontiers} in {Education} {Conference} ({FIE})},
  pages 1--8, October 2020.
\newblock ISSN: 2377-634X.
\newblock \href {https://doi.org/10.1109/FIE44824.2020.9274144}
  {\path{doi:10.1109/FIE44824.2020.9274144}}.

\bibitem{spring_review_2021}
Jonathan~M. Spring and Phyllis Illari.
\newblock Review of {Human} {Decision}-making during {Computer} {Security}
  {Incident} {Analysis}.
\newblock {\em Digital Threats: Research and Practice}, 2(2):1--47, June 2021.
\newblock URL: \url{https://dl.acm.org/doi/10.1145/3427787}, \href
  {https://doi.org/10.1145/3427787} {\path{doi:10.1145/3427787}}.

\bibitem{stoker_building_2021}
Geoff Stoker, Ulku Clark, Manoj Vanajakumari, and William Wetherill.
\newblock Building a {Cybersecurity} {Apprenticeship} {Program}:
  {Early}-{Stage} {Success} and {Some} {Lessons} {Learned}.
\newblock page~10, 2021.

\bibitem{sundaramurthy_tale_2014}
Sathya~Chandran Sundaramurthy, Jacob Case, Tony Truong, Loai Zomlot, and Marcel
  Hoffmann.
\newblock A {Tale} of {Three} {Security} {Operation} {Centers}.
\newblock In {\em Proceedings of the 2014 {ACM} {Workshop} on {Security}
  {Information} {Workers} - {SIW} '14}, pages 43--50, Scottsdale, Arizona, USA,
  2014. ACM Press.
\newblock \href {https://doi.org/10.1145/2663887.2663904}
  {\path{doi:10.1145/2663887.2663904}}.

\bibitem{sundaramurthy_turning_2016}
Sathya~Chandran Sundaramurthy, John McHugh, Xinming Ou, Michael Wesch,
  Alexandru~G. Bardas, and S.~Raj Rajagopalan.
\newblock Turning contradictions into innovations or: How we learned to stop
  whining and improve security operations.
\newblock In {\em Twelfth Symposium on Usable Privacy and Security (SOUPS
  2016)}, pages 237--251, Denver, CO, June 2016. USENIX Association.
\newblock URL:
  \url{https://www.usenix.org/conference/soups2016/technical-sessions/presentation/sundaramurthy}.

\bibitem{sundaramurthy_anthropological_2014}
Sathya~Chandran Sundaramurthy, John McHugh, Xinming~Simon Ou, S.~Raj
  Rajagopalan, and Michael Wesch.
\newblock An {Anthropological} {Approach} to {Studying} {CSIRTs}.
\newblock {\em IEEE Security Privacy}, 12(5):52--60, September 2014.
\newblock Conference Name: IEEE Security Privacy.
\newblock \href {https://doi.org/10.1109/MSP.2014.84}
  {\path{doi:10.1109/MSP.2014.84}}.

\bibitem{symonds_innovating_2017}
R~Symonds.
\newblock Innovating the {Prioritization} of {Cyber} {Defense}.
\newblock {\em Journal of Information Warfare}, 16(2):12--18, 2017.
\newblock Publisher: Peregrine Technical Solutions.
\newblock URL: \url{http://www.jstor.org/stable/26502753}.

\bibitem{trent_modelling_2019}
Stoney Trent, Robert~R. Hoffman, David Merritt, and Sarah Smith.
\newblock Modelling the {Cognitive} {Work} of {Cyber} {Protection} {Teams}.
\newblock {\em The Cyber Defense Review}, 4(1):125--136, 2019.
\newblock Publisher: Army Cyber Institute.

\bibitem{kleij_computer_2017}
Rick Van~der Kleij, Geert Kleinhuis, and Heather Young.
\newblock Computer security incident response team effectiveness: A needs
  assessment.
\newblock {\em Frontiers in Psychology}, 8, 2017.
\newblock \href {https://doi.org/10.3389/fpsyg.2017.02179}
  {\path{doi:10.3389/fpsyg.2017.02179}}.

\bibitem{van_os_tahiti_2018}
Rob van Os, Marcus Bakker, Ruben Bouman, Martijn~Docters van Leeuwen, Marco
  van~der Kraan, and Wesley Mentges.
\newblock {TaHiTI}: a threat hunting methodology, December 2018.
\newblock URL:
  \url{https://www.betaalvereniging.nl/wp-content/uploads/TaHiTI-Threat-Hunting-Methodology-whitepaper.pdf}.

\bibitem{wafula_carve_2019}
Kevin Wafula and Yong Wang.
\newblock {CARVE}: {A} {Scientific} {Method}-{Based} {Threat} {Hunting}
  {Hypothesis} {Development} {Model}.
\newblock In {\em 2019 {IEEE} {International} {Conference} on {Electro}
  {Information} {Technology} ({EIT})}, pages 1--6, May 2019.
\newblock ISSN: 2154-0373.
\newblock \href {https://doi.org/10.1109/EIT.2019.8833792}
  {\path{doi:10.1109/EIT.2019.8833792}}.

\bibitem{wei_deephunter_2021}
Renzheng Wei, Lijun Cai, Aimin Yu, and Dan Meng.
\newblock {DeepHunter}: {A} {Graph} {Neural} {Network} {Based} {Approach} for
  {Robust} {Cyber} {Threat} {Hunting}, April 2021.
\newblock arXiv:2104.09806 [cs].
\newblock URL: \url{http://arxiv.org/abs/2104.09806}.

\bibitem{furnell_preparation_2010}
Rodrigo Werlinger, Kasia Muldner, Kirstie Hawkey, and Konstantin Beznosov.
\newblock Preparation, detection, and analysis: the diagnostic work of {IT}
  security incident response.
\newblock {\em Information Management \& Computer Security}, 18(1):26--42,
  March 2010.
\newblock \href {https://doi.org/10.1108/09685221011035241}
  {\path{doi:10.1108/09685221011035241}}.

\bibitem{white_btfm_2017}
Alan White and Ben Clark.
\newblock {\em {BTFM}: blue team field manual}.
\newblock Alan White, United States, version 1, rel 2 edition, 2017.

\bibitem{whitehead_ukraine_2017}
David~E Whitehead, Kevin Owens, Dennis Gammel, and Jess Smith.
\newblock Ukraine cyber-induced power outage: {Analysis} and practical
  mitigation strategies.
\newblock In {\em 2017 70th {Annual} {Conference} for {Protective} {Relay}
  {Engineers} ({CPRE})}, pages 1--8. IEEE, 2017.

\bibitem{wilson_evaluation_2015}
John~R. Wilson and Sarah Sharples, editors.
\newblock {\em Evaluation of {Human} {Work}}.
\newblock CRC Press, Boca Raton, 4 edition, April 2015.
\newblock \href {https://doi.org/10.1201/b18362} {\path{doi:10.1201/b18362}}.

\end{thebibliography}


\ifTECHREPORT
\clearpage
\appendix

\section{Summary of Appendices} \label{sec:Appendix-Summary}

\begin{itemize}
    \item \cref{sec:Appendix-Summary}: This summary.
    \item \cref{sec:Appendix-InterviewProtocol}: The interview protocol.
    \item \cref{sec:Appendix-SaturationCharts}: Saturation charts showing that our N=11 interviews saturated the coding space under consideration.
    \item \cref{sec:Appendix-JobRoleAnalysis}: Discussion of the effect of a subject's rank on our thematic analysis.
    \item \cref{sec:Appendix-Codebooks}: The codebooks used in our analysis, with illustrative quotes mapped to each code.
    \item \cref{sec:Appendix-THFrameworks}: To keep this manuscript self-contained, we include details about cybersecurity frameworks --- Cyber Kill Chain and the Pyramid of Pain.
    \item \cref{sec:Appendix-RelatedProcesses}: To keep this manuscript self-contained, we include details about related TH processes --- the \tahiti process diagram, and the Trent \etal cognitive model from which we extracted another TH process.
    \item \cref{sec:Appendix-MethodologyDetails}: Additional details of our methodology and results, notably a detailed TH process diagram (\cref{fig:DetailedInducedDiagram}) refining the high-level version presented earlier (\cref{fig:CoarseInducedDiagram}).
\end{itemize}


\section{Interview Protocol}
\label{sec:Appendix-InterviewProtocol}

This interview guide is organized into headings that identify the main themes or lines of inquiry, under which individual questions to be asked are given. The main questions are listed first, with possible follow-on probes nested underneath. Due to the nature of a semi-structured interview, the questions may not be asked exactly as written or in the same sequence, but may be adjusted depending on how the conversation flows. The intention is that they serve mainly as a reminder or checklist for the interviewer. 

A ``*'' indicates a question added either after the 2 pilot studies or after a theme emerged from the first few interviews. 


\subsection{Introduction (10 mins)}
After 2 administrative reminders, this portion of the interview will be focused on discerning the history of the interviewee and enumerating the teams on which they were a member. If they were threat hunters for multiple organizations then the other sets of questions may be able to be asked about multiple organizations with the understanding that some of the information may be outdated. 

\subsubsection{Reminders:}
\begin{itemize}
 \item This interview is being conducted over an insecure medium. Please keep all discussion limited to only Unclassified information and refrain from providing customer details that should remain confidential. 
 \item This interview is being recorded. 
 \item Do you have any questions about this interview's classification policy or any general questions about this interview, the study, the data being collected or the researchers?
\end{itemize}

\subsubsection{Questions:}
\begin{itemize}
 \item How long have you been on your current threat hunt team?
 \begin{itemize}
 \item How many missions have you been engaged on?
 \item How many of your missions successfully identified adversary activity? 
 \end{itemize}
 \item Have you worked on any threat hunt teams previously?
 \begin{itemize}
 \item Which ones? For how long?
 \item How many missions were you deployed on with your previous team?
 \item How many of those missions successfully identified adversary activity? 
 \end{itemize}
\end{itemize}

\subsection{Threat hunting processes (30 mins)} 
This portion of the interview focuses on the process used by an agency and how that process came about, how effective it is, how it could be improved, etc. This will primarily contain questions relevant to research question 1. 

\subsubsection{Questions:}
\begin{itemize}
 \item Could you please draw a diagram of the current threat hunting process that is followed by your team and explain the diagram?
 \begin{itemize}
 \item Is this process standardized across all teams in your agency? If no, why not?
 \item How often is the process modified?
 \item What causes the process to be modified?
 \item How strictly is the process followed? 
 \item Do you believe this is too strict or to lose? Why?
 \item What do you find to be the most problematic parts of this process? Why?
 \end{itemize}
 \item What models are incorporated in your process? An example model would be the OODA loop. 
 \item In what way do you incorporate the following models into your process:
 \begin{itemize}
 \item Kill chain Model?
 \item Pyramid of Pain?
 \item MITRE \attack?
 \item Hypothesis creating and checking?
 \begin{itemize}
 \item Can you give us an example of a hypothesis? 
 \item How are they documented and shared with the team?
 \end{itemize}
 \end{itemize}
 \item How is the process tracked or process documentation accessed when on engagement?
 \begin{itemize}
 \item How often is it referenced?
 \item How granular is the process documentation?
 \item Why that level of granularity?
 \item Do you believe it is too specific or vague? Why?
 \end{itemize}
 \item How was the current process developed? 
 \begin{itemize}
 \item Were any alternative processes or frameworks considered?
 \item Who was given the opportunity to provide input in the development and institution of the process?
 \end{itemize}
 \item Can you think of a mission where the TH process was not as helpful as you would have liked or even counter productive? (Question should be sent ahead of time so the subject have time to remember an event)
 \begin{itemize}
 \item Please describe the event to without giving us details about the customer. 
 \item How long did it take before the process deficiencies were made obvious?
 \item What was the shortcoming? 
 \item What changes to the process would help alleviate this (these) shortcoming(s)?
 \item How common are experiences like this?
 \item Did the experience level of the team contribute to the issue?
 \end{itemize}
 \item Can you think of a second mission where the TH process was not as helpful as you would have liked or even counter productive? (Question should be sent ahead of time so the subject have time to remember an event)
 \begin{itemize}
 \item Please describe the event to without giving us details about the customer. 
 \item How long did it take before the process deficiencies were made obvious?
 \item What was the shortcoming? 
 \item What changes to the process would help alleviate this (these) shortcoming(s)?
 \item How common are experiences like this?
 \item Did the experience level of the team contribute to the issue?
 \end{itemize}
 \item How much of the process is automated?
 \begin{itemize}
 \item Who automated it?
 \item How much more of it can be automated?
 \item Why hasn’t it been?
 \end{itemize}
 \item When adversary activity is typically found on a Threat Hunt, how is it typically found?
 \begin{itemize}
 \item How experienced is the member that typically finds it?
 \end{itemize}
 \item Roughly how often is adversary activity found:
 \begin{itemize}
 \item With a sensor alert?
 \item With a tip from another team conducting an Incident Response mission?
 \item With a tip from intelligence sources?
 \item Another way?
 \begin{itemize}
 \item When the activity is discovered another way, is there a common way or location it is discovered?
 \item How often is the person who discovers the activity another way an inexperienced member?
 \end{itemize}
 \end{itemize}
 \item Does your process include a mission plan?*
 \begin{itemize}
 \item Is it helpful?*
 \item Does it ever get in the way of operations?*
 \item What would make it more helpful?*
 \end{itemize}
\end{itemize}

\subsection{Integration of new members (20 mins)}
This portion of the interview focuses on how newer members are affected by a change in process and what improvements could be made to onboard new members more efficiently. These questions correspond primarily to research questions 2 and 3. 

\subsubsection{Questions:}
\begin{itemize}
 \item How long does it take new members of the team to independently reason through the current process without help? 
 \begin{itemize}
 \item Why does it take this long?
 \item Are there variables that speed this up or slow it down?
 \item Does this change if the new member has threat hunt experience?
 \item Does this change if the new member has other experiences in their background? What experiences are helpful?
 \item How long does it take new members to become a net positive for the team when on engagement?
 \item Does this happen at the point they can reason through the current process independently or at a different time? (This question and follow-up question were removed after 5 interviews)
 \begin{itemize}
 \item Why do you think this is?
 \end{itemize} 
 \end{itemize}
 \item What is a good measure of a member's expertise?*
 \begin{itemize}
 \item Experience?*
 \item Certifications?*
 \item Personality?*
 \item Willingness to put in own time?*
 \item Is this different for potential?*
 \end{itemize}
 \item Have you observed any changes to the process that allow for a faster integration of new members?
 \begin{itemize}
 \item What were they?
 \item Where these changes made permanent? 
 \item If no, why not?
 \item To what extent have other team leads observed a similar phenomenon?
 \item Would you have any other recommendations?*
 \end{itemize}
 \item Have you observed any changes that slow down the integration of new members?
 \begin{itemize}
 \item What were they?
 \item Where these changes made permanent?
 \item To what extant have other team leads observed a similar phenomenon?
 \end{itemize}
 \item Before we close, are there any additional questions that I should have asked but didn't?
 \item Is there anything else I should know about threat hunting processes?
\end{itemize}

\subsection{Closing (2 mins)}
We thank the participant and bring the interview to a close. 

\section{Saturation Charts} \label{sec:Appendix-SaturationCharts}

This appendix contains saturation charts from our thematic coding (used for RQ1 and RQ2).

\cref{sat_fig} shows the cumulative number of unique codes plotted by subject.

\begin{figure}[htb]
\includegraphics[width=\columnwidth]{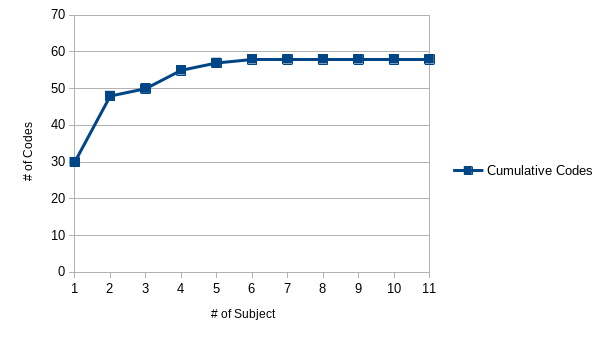}
\caption{
  Cumulative unique codes by subject.
  The graph indicates saturation was achieved after 7 subjects.
  }
\label{sat_fig}
\end{figure}

\cref{code_fig} shows the number of unique codes per subject.

\begin{figure}[htb]
\includegraphics[width=\columnwidth]{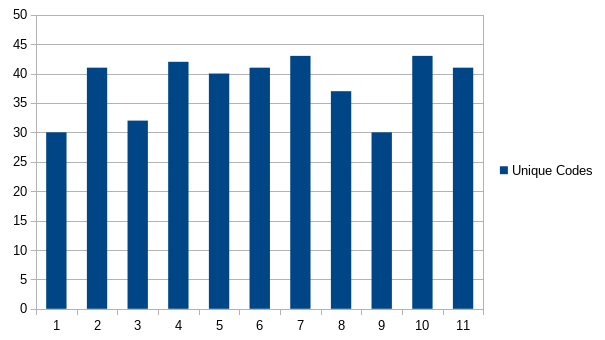}
\caption{
  Unique codes observed per subject.
  The graph indicates that re-ordering the saturation chart (\cref{sat_fig}) would not affect the saturation curve.
  }
\label{code_fig}
\end{figure}

\section{Analysis of Themes by Job Role} \label{sec:Appendix-JobRoleAnalysis}

As noted in~\cref{table:SubjectDemographics}, we stratified our sampling by the subject's job role.
Our sample included
  4 subjects in organizational leadership roles,
  4 in team leader roles,
  and 
  3 in analyst roles.
We wondered whether our subjects would report different processes or experiences based on their role on a TH team.

Our subjective impression was that the team leads tended to have the most awareness of what the TH process entailed --- one author conducted all interviews and formed this opinion. 
We speculate this could be because they are the `enforcers' of the process with the analysts looking primarily at their tasking and the leadership not concerned with the details of how the hunt is being conducted. 

To assess this more quantitatively, we examined the frequency at which themes came up by job role.
\cref{fig:ThemesJobRoleAnalysis} shows the result applied to the first 20 codes in our codebook, alphabetically.
It does not appear that our topical references are strongly impacted by job roles.
We did not, however, analyze whether job role influenced the tenor of our subjects' remarks in each theme.


\begin{figure*}[ht]
\centering
\includegraphics[scale=0.85]{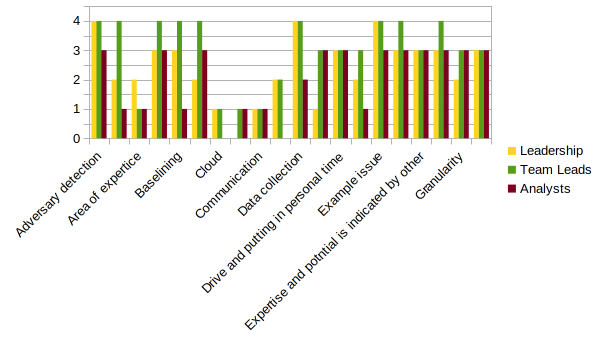}
\caption{
  Frequency at which thematic codes were mentioned by subjects, grouped by job role.
  The chart depicts the first 20 codes alphabetically.
  For visual clarity, the x-axis names only half of the codes.
  We observe no systematic variation by job role.
  }
\label{fig:ThemesJobRoleAnalysis}
\end{figure*}

\section{Sample Codes} \label{sec:Appendix-Codebooks}

This appendix contains illustrative excerpts from our codebooks.
The full codebooks will be available as an artifact accompanying the published paper.

\subsection{Process Coding to \tahiti} \label{sec:Appendix-Codebooks-Tahiti}

We provide three codes from the codebook we used for process coding against \tahiti in~\cref{table:TahitiCodebook}.
We developed this codebook by reviewing the \tahiti model, having one code per node in their process model, and providing definitions for each of these nodes.

\begin{table*}
    \centering
    \caption{
    Excerpt from the codebook we used for process coding against \tahiti~\cite{van_os_tahiti_2018}. 
    }
    \label{table:TahitiCodebook}
    \begin{tabularx}{\textwidth}{l X p{0.5\textwidth}}
        \toprule
        \textbf{Code} & \textbf{Definition} & \textbf{Interview quotes mapped to this code} \\
        \toprule
        Determine Hypothesis & Explicitly or implicitly describing the act of creating a hypothesis & 
        \begin{itemize}[topsep=0pt, partopsep=0pt]
            \item ``That would be an example of specific intelligence where you now have something that you can say, okay, we’re going in with the hypothesis that we are compromised specifically. We’re gonna start with this system and we’re gonna hunt from there.''
            \item ``We can’t hunt for everything over the scope of our engagement so really we’re just trying to come up with the best hypothesis we can, to then go hunt for that activity.''
        \end{itemize} \\
        Retrieve Data & Having to do with the collection of logs and network or host based artifacts. & 
        \begin{itemize}[topsep=0pt, partopsep=0pt]
            \item ``We try to deploy sensors, depending on the timeliness of it. In this case deploying as soon as possible -- more data is better, obviously.''
            \item ``We went on site for two weeks, gathered customer data for those two weeks, using [tool] and all of the other tools that we had.''
            \item ``We place sensors. We’ll then let those run and collect metadata and network data from the network that we’re concerned about.''
        \end{itemize} \\
        Analyze Data & Having to do with a TH team member examining the collected logs and artifacts. & 
        \begin{itemize}[topsep=0pt, partopsep=0pt]
            \item ``Days three to probably 12 will actually comb through all of the data. We will split off into network versus host based analytics. We’ll have our network analysts go through all the network data and our host analysts go through host data.''
            \item ``The team uses the analytical repository and based on the mission plan, they start running those analytics.''
            \item ``If anything pops positive from there, we'll move on to our anomaly-based, analysis.''
        \end{itemize} \\
        \bottomrule
    \end{tabularx}
\end{table*}

\begin{table*}
    \centering
    \caption{
    Excerpt from the codebook we used for process coding against Trent \etal~\cite{trent_modelling_2019}. 
    }
    \label{table:TrentCodebook}
    \begin{tabularx}{\textwidth}{l X p{0.5\textwidth}}
        \toprule
        \textbf{Code} & \textbf{Definition} & \textbf{Interview quotes mapped to this code} \\
        \toprule
        Deploy or Adjust Sensors & Having to do with the setup or configuration of any type of data sources (Network or Host based). & 
        \begin{itemize}[topsep=0pt, partopsep=0pt]
            \item ``So, at least two weeks out from mission start date, we place sensors.''
            \item ``and we're like, okay, let's try to pre-deploy the kit several weeks ahead of time.''
            \item ``You might need to attach some additional sensors though, to go look for what you wanna look for. ''
        \end{itemize} \\
        Network Analysis & Analysis occuring on network data, including general analysis where one of the data sources is network related (e.g. netflow, pcap, firewall log, etc.). & 
        \begin{itemize}[topsep=0pt, partopsep=0pt]
            \item ``We will split off into network versus host based analytics. So we'll have our network analysts go through all the network data and our host analysts...
            ''
            \item ``So using the network logs that we have, um, and then context with the local [network administrators], it trying to figure out, you know, where their domain controllers are, where their main servers are.''
            \item ``Then we begin our indicator-based analysis.''
        \end{itemize} \\
        Terrain Characterization & Baselining or seeking to understand data flow, network flow or endpoints on the network. & 
        \begin{itemize}[topsep=0pt, partopsep=0pt]
            \item ``We got to figure out what kind of environment we're going to be working in, how many endpoints, how many users, inventory, all the how much data flowing through their networks, stuff like that.''
            \item ``You've got people mapping out the network, asking questions to the network owner, getting our baseline of what are we looking at? What's your critical terrain? What's your, where's your DMZs? Where is any possible PII, sensitive information? Anything like that.''
            \item ``We want to see, again, an overall picture of, um, the servers, the works stations, network infrastructure, um, key assets that they have that are most concerned about as well as, um, if they have any, uh, infrastructure, should they believe is compromised, non-compromised, et cetera.''
        \end{itemize} \\
        \bottomrule
    \end{tabularx}
\end{table*}

\subsection{Process Coding to Trent} \label{sec:Appendix-Codebooks-Trent}

We followed a similar process to develop a codebook for the Trent model.
Note that the Trent model covers all activities of Cyber Protection Teams (CPT), and CPTs conduct cybersecurity activities beyond the scope of a threat hunt (\eg incident response).
Therefore, in developing the codebook for process coding against Trent, we omitted activities unrelated to threat hunt.
An excerpt from the resulting codebook is in~\cref{table:TrentCodebook}.

\subsection{Thematic Coding}

We provide example codes from our thematic codebook (used for RQ1 and RQ2).
See~\cref{table:ThematicCodebook}.

\begin{table*}
    \centering
    \caption{Excerpt from our codebook for thematic coding. Themes were iteratively induced from the data.}
    \label{table:ThematicCodebook}
    \begin{tabularx}{\textwidth}{l X p{0.5\textwidth}}
        \toprule
        \textbf{Code} & \textbf{Definition} & \textbf{Examples from interviews} \\
        \toprule
        Intelligence & Pertaining to the team's intelligence component, intel staff, IOCs, open source intel or other intel sources & 
        \begin{itemize}[topsep=0pt, partopsep=0pt]
            \item ``This last one’s the most generic, but like in the absence of everything else, you can say current attacker trends.''
            \item ``I would say the alerting to potential malicious actors is built into the Intel process and then the team can look up their tactics.''
            \item ``Intel is always hard because sometimes the Intel people don’t necessarily understand what’s going on. A lot of traditional Intel people struggle when it comes to cyber space operations.''
        \end{itemize} \\
        Automation & Having to do with automated TH tasks ---- can be tactical, administrative or other & 
        \begin{itemize}[topsep=0pt, partopsep=0pt]
            \item ``The issue also with automation is like, like I said, our hunts are only two weeks long, so a lot of, some of the cool automation tools we want to use that we have in our tool bucket right now would take too long.''
            \item ``None of it’s automated as of yet. I know it’s in the pipeline but I don’t see it happening anytime soon.''
            \item ``I would say that you should not have automated things available to the teams as part of the checklist. Each team would need to build their own automations, for their own style and technique. If you provide people automated analysis, they will run that and it will become a crutch that they will not do analysis on their own.''
        \end{itemize} \\
        Mission Plan & Specifically mentions mission plan or a similar document outlining objectives or hypothesis & 
        \begin{itemize}[topsep=0pt, partopsep=0pt]
            \item ``The mission plan will stay with our [organization], the parent [organization] and the mission partner. Other than that, we have to anonymize everything to not attributable status if we want to share outside of those.''
            \item ``In the mission plan, we’ll include any intelligence that we have based on the scoping documentation, any previous incidents that we discovered during talking to the mission partner.''
            \item ``With the knowledge of the network, the knowledge of any Intel that we have, we create a mission plan.''
        \end{itemize} \\
        \bottomrule
    \end{tabularx}
\end{table*}

\begin{table*}
    \centering
    \caption{
    \new{Our full codebook for thematic coding. Themes, sub-themes, and codes were iteratively induced from the data.}
    }
    \scriptsize
    \label{table:FullThematicCodebook}
    \begin{tabularx}{\textwidth}{l l l X}
        \toprule
        \textbf{Theme} & \textbf{Topic} & \textbf{Code} & \textbf{Definition}\\
        \toprule
        \multirow{40}{*}{\rotatebox[origin=c]{90}{\footnotesize PROCESS}} & Process Phases & Adversary detection & Related to how adversaries are caught by a TH team and how often\\ 
         &  & Baselining & Specifically mentions baselining or describes a similar process\\ 
         &  & Data collection & Related to data collection challenges, solutions, and current practice \\ 
         &  & Discussion with customer & Related to talking to a customer/mission partner\\ 
         &  & Intelligence  & Related to the team's intelligence component, intel staff, IOCs, open-source intel, or other intel sources\\ 
         &  & IR team called & Related to the IR team, an IR mission or an adversary being found\\ 
         &  & Mission Plan & Specifically mentions mission plan or a similar document outlining objectives or hypothesis\\ 
         &  & Objectives & Describes the importance of or role of goals in the threat hunting process or in documentation\\ 
         &  & Reporting & Related to the final product or closing meeting of a hunt\\ 
         &  & Sensor deployment & Anything mentioning sensors \\ 
         \cmidrule{2-4}
         & Process Issues & Example issue & Describes an issue that the subject observed on a mission\\
         &  & Not doing ad-hoc hunting & Reasons not to do ad-hoc hunting\\ 
         &  & Personnel issues & A catch-all for personnel issues\\ 
         &  & Probability of issues & Related to the frequency of issues and certain issue types\\ 
         &  & Timeline for realizing deficiency & Related to how quickly a team, member or leader knew that the process was deficient\\ 
         \cmidrule{2-4}
         & TH Models & Hypothesis Method & Specifically uses the word hypothesis or describes a similar process\\ 
         &  & Mitre ATT\&CK & Specifically mentions Mitre and/or ATT\&CK\\ 
         &  & Other models & Any other model specifically mentioned (PoP, Kill Chain, TaHiTI)\\ 
         \cmidrule{2-4}
         & Process Stability & Enforcement & Related to the enforcement of certain process tasks\\ 
         &  & Flexibility & Describes how often a non-codified change is made to the process or how often the process is changed/ignored onsite\\ 
         &  & Granularity & Related to how specific a process needs to be \\ 
         &  & Process change & Describes how often a codified change is made to the process\\ 
         &  & Process Creation & Related to the genesis of the process at the subject's organization\\ 
         &  & Process Referencing & Relevant to the the way and frequency a team members looks at the process (or a process related document) while on site\\ 
         &  & Scope of Documentation & Documentation done by Analysts\\ 
         &  & Standardization of Process & Describes how similar the process is across team internal and external to the organization of the subject\\ 
         &  & Task list & Related to a list complied of task to achieve and how that fits into the process\\ 
         \cmidrule{2-4}
         & Other & Cloud & Related to any cloud system\\ 
         &  & Collaterals & Related to TH members having secondary jobs \\ 
         &  & Iterative processes & References cycles or parts of the process that are repeated\\ 
         &  & Recurrence and parallel task & Related to simultaneously occurring or continuously occurring tasks\\ 
         &  & Time & Related to process timing\\ 
         \cmidrule{2-4}
         & Tools & Automation & Related to automated TH tasks --- can be tactical, administrative or other\\ 
         &  & Tools & Related to the tools being used by the TH members\\ 
         &  & Tracking system & Related to how process tracking was done while team is on mission\\ 
        \toprule
        \multirow{8}{*}{\rotatebox[origin=c]{90}{\footnotesize NEW MEMBERS}} & Processes & Process hurts new members & Observed process changes that \textit{hurt} new members \\ 
         &  & Process help new members & Observed process changes that \textit{help} new members\\ 
         \cmidrule{2-4}
         & Other & Apprenticeship & Descriptions of Apprenticeship\\ 
         &  & Helpful backgrounds & Discussing what previous experiences assist time to integration\\ 
         &  & No Improvement & Related to the belief that nothing will improve analysts who are not already performing well\\ 
         &  & Other missions & Related to sending new members on non-hunt missions\\ 
         &  & Time to minimum capability & Related to the amount of time it takes a new member to become an expert\\ 
        \toprule
        \multirow{14}{*}{\rotatebox[origin=c]{90}{\footnotesize EXPERTISE}} & Expertise Signifiers & Area of expertise & Related to being outstanding due to a specific area of expertise or certain focus\\ 
         &  & Certificates & Related to being outstanding due to certifications\\ 
         &  & Communication & Related to being outstanding due to communication skills\\ 
         &  & Curiosity & Related to being outstanding due to being curious\\ 
         &  & Drive/putting in personal time & Related to being outstanding due to putting in time outside of work or other indicators of drive\\ 
         &  & Expertise is indicated by other & Related to being outstanding due to something not yet coded\\ 
         &  & JQRs & Related to the Job Qualification Requirement (JQR) document or equivalent document\\ 
         &  & Mindset & Related to the importance of mindset\\ 
         &  & Number of missions/experience & Related to being outstanding due to a high number of missions\\ 
         &  & Personality & Related to being outstanding due to a certain personality\\ 
         &  & Teamwork & Related to members working together\\ 
         &  & Training & Related to being outstanding due to training or descriptions of training and its issues\\
         \cmidrule{2-4}
         & Other & Pre-mature ``investigations'' & Related to analysts not fully investigating their assigned task\\ 
         &  & Rabbit holes & Related to one member becoming too engrossed in an investigation\\ 
         &  & Task separation & Related to separating analysts tasking by expertise\\ 
        \bottomrule
    \end{tabularx}
\end{table*}

\section{Existing TH Frameworks} \label{sec:Appendix-THFrameworks}

This appendix contains additional relevant Threat Hunt materials.

\cref{fig:RelatedFrameworksAndModels} contains two common Threat Hunt frameworks: the Lockheed Martin Cyber Kill Chain and Bianco's Pyramid of Pain.
The Mitre \attack Framework can be viewed at \url{https://attack.mitre.org}.

\begin{figure*}[b]
    \centering
    \begin{subfigure}[b]{\textwidth}
        \centering
        \includegraphics[height=3cm]{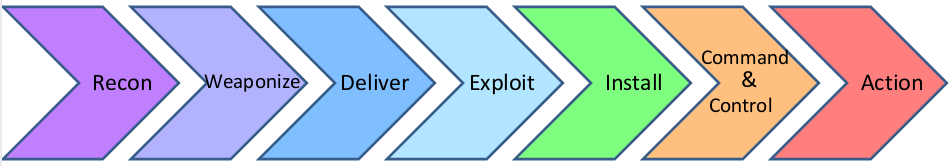}
        \caption{
        Lockheed Martin's Cyber Kill Chain~\cite{noauthor_kill_2014}.
        }
        \label{fig:KillChain}
    \end{subfigure}
    \bigskip
    \begin{subfigure}[b]{\textwidth}
        \centering
        \includegraphics[height=5cm]{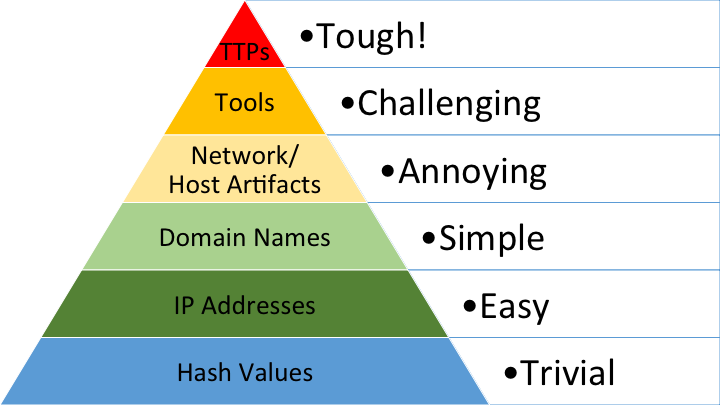}
        \caption{Bianco's Pyramid of Pain~\cite{davidjbianco_enterprise_2013}.}
        \label{fig:PyramidOfPain}
    \end{subfigure}
    \caption{Two example Cybersecurity Frameworks. TTPs are Tactic, Technique, and Procedures }
    \label{fig:RelatedFrameworksAndModels}
\end{figure*}

\section{TH Diagrams from Prior Work} 
\label{sec:Appendix-RelatedProcesses}

This appendix contains diagrams from related work.

\subsection{The \tahiti TH Process}

\cref{fig:TahitiProcessModel} is the threat hunt process used by four Dutch financial organizations, as described by~\cite{van_os_tahiti_2018}.

\begin{figure*}[b]
\centering
\includegraphics[scale=0.45]{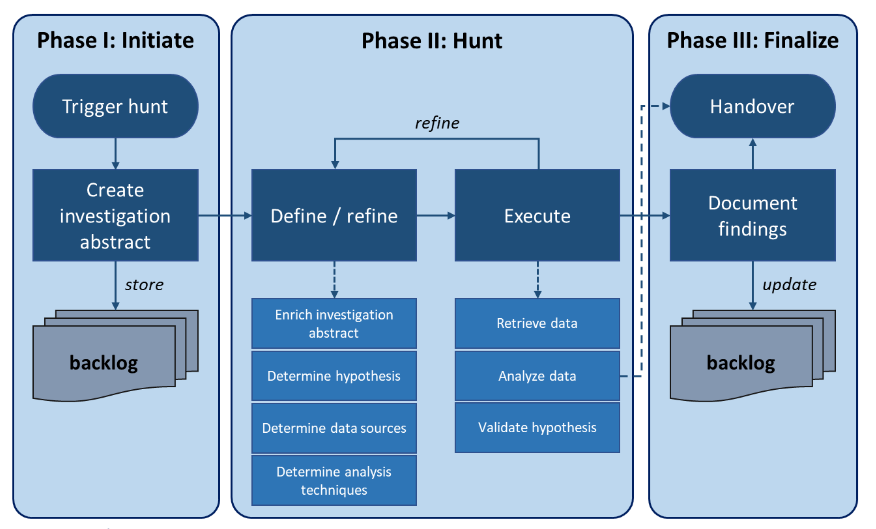}
\caption{The \tahiti process from van Os \etal~\cite{van_os_tahiti_2018}.}
\label{fig:TahitiProcessModel}
\end{figure*}

\subsection{The Trent Model}

\cref{fig:TrentCPTWorkModel} is a cognitive work model of US military Cyber Protection Teams (CPTs), as developed by Trent \etal~\cite{trent_modelling_2019}. 

\begin{figure*}[b]
\centering
\includegraphics[scale=0.5]{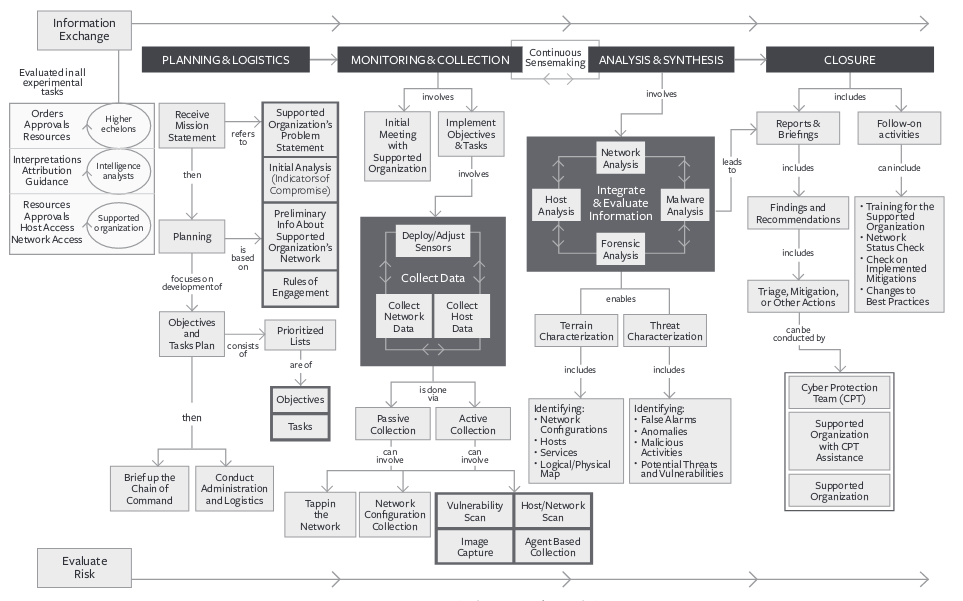}
\caption{Detailed CPT work model developed by Trent \etal~\cite{trent_modelling_2019}.}
\label{fig:TrentCPTWorkModel}
\end{figure*}
 
\subsection{Bianco's TH Maturity Framework}

\cref{thmm} provides Bianco's maturity model for Threat Hunt teams~\cite{bianco_simple_nodate}.

\begin{figure*}[b]
\includegraphics[width=\textwidth]{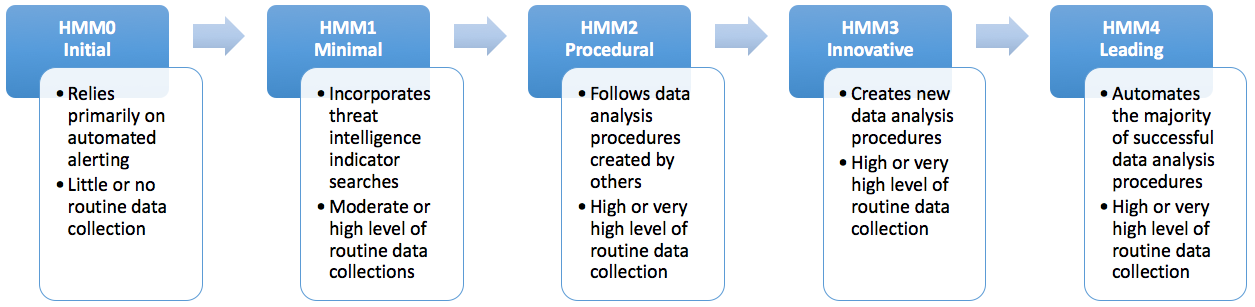}
\caption{Bianco's Threat Hunting Maturity Model~\cite{bianco_simple_nodate}.}
\label{thmm}
\end{figure*}






\section{Additional methodology and results} \label{sec:Appendix-MethodologyDetails}

\begin{table}[]
\caption{
  Subjects' experience and missions. * indicates that subject interview was a pilot interview.
}
\centering
{
\small
\renewcommand{\arraystretch}{0.8}
\begin{tabular}{ccc}
\toprule
\textbf{Subject} & \textbf{Years of Experience} & \textbf{Number of TH Missions} \\ \midrule
1* & 5 & 10 \\
2* & 2 & 6 \\
3 & 6 & 24 \\
4 & $<1$ & 1 \\
5 & 1 & 3 \\
6 & 1 & 0 \\
7 & 4 & 2 \\
8 & 7 & 5 \\
9 & 1 & 1 \\
10 & $<1$ & 1 \\
11 & 2 & 2 \\ \bottomrule
\end{tabular}
}
\label{subs_vs_exp_mis}
\end{table}

\subsection{Detailed definition of TH personnel types} \label{mapping}

\begin{itemize}
\item \emph{Leadership} personnel deal with a hunt team's strategic concerns. Leaders did not deploy with the team. Job titles of personnel in this category include analyst's managers, officers in command, and executive, operations, and analysis officers.
\item \emph{Team Leads} are the most senior member that deploys with a hunt team, and for some teams, there is also a deputy in this category.
\item \emph{Analysts} perform the analytic tasks associated with hunting. Analysts deploy with the hunt team and sometimes perform setup and maintenance the TH equipment while deployed.
\end{itemize}

\subsection{Details of evolution of Interview protocol} \label{protocol}

After the pilot studies, we modified the original interview protocol when we observed phenomena not covered by the protocol.
For example, early participants referenced a document called a ``Mission Plan''; we added a question about its utility. 

The changes after the pilot were:
\begin{itemize}
\item 2 follow-up questions were removed
\item 2 questions were added
\item 9 sample follow-up questions were added
\end{itemize}

These changes are annotated in~\cref{sec:Appendix-InterviewProtocol}.

\subsection{Results on induced process model}

To simplify the presentation of this work, the TH process model presented in~\cref{fig:CoarseInducedDiagram} is a high-level version.
A detailed full-page version is given in~\cref{fig:DetailedInducedDiagram}.
This version includes details about what occurs in each high-level stage of~\cref{fig:CoarseInducedDiagram}, as well as annotation suggesting possible points at which TH teams could incorporate existing TH frameworks (our analysis).

\begin{figure*}[h!]
\includegraphics[width=0.93\textheight,height=0.93\textwidth,angle=270]{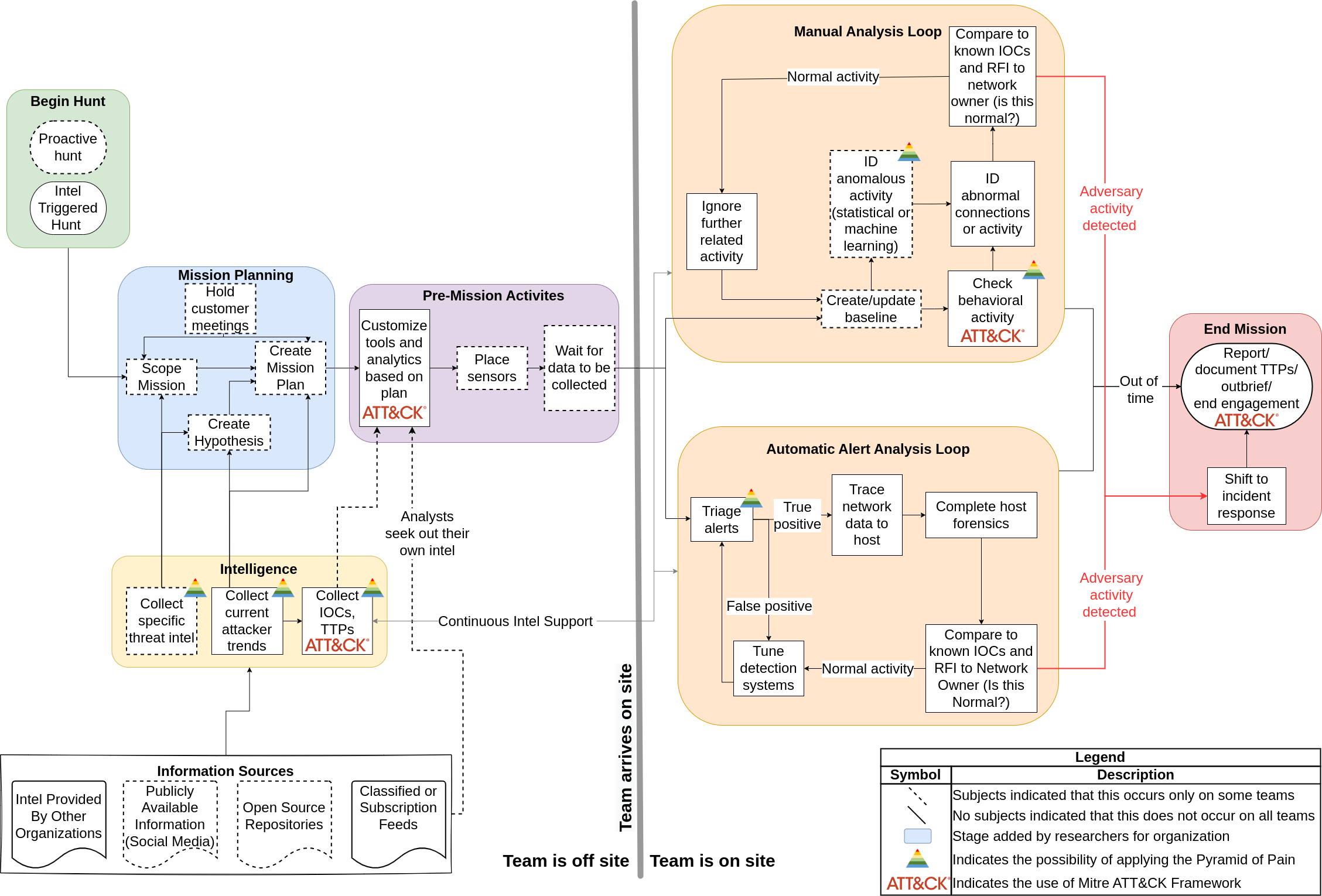}
\caption{
  Diagram synthesized from subject's hand-drawn diagrams. 
  If an activity was: (1) in multiple diagrams or (2) one diagram and referenced by another subject, it was added to this diagram. 
  Activities were grouped by researchers into stages. 
  }
\label{fig:DetailedInducedDiagram}
\end{figure*}

\subsection{Results on process consistency across organizations:}
\label{standard_rq1}

Across all three organizations, subjects claimed everyone in their organization used the same process.
2 subjects stated the process would look similar within the DHS but beyond their agency. 
This claim was also paired with statements indicating the process's flexibility. 
One subject discussed a spreadsheet for process tracking,
but that the team lead decided what went on it. 
Another subject said the team lead could omit the spreadsheet entirely.  
A team lead said: \inlinequote{there is a checklist that kind of went over the steps I alluded to but I would [say, in] my own personal opinion, the steps are useless and nobody actually followed them...the checklist was there for someone to ensure minimum quality standards.}
This somewhat contradicted that subject's earlier statement that their organization's process is standardized. 
Contradictions like this indicate that processes being used within an organization may vary team to team.

\subsection{Results on subjects' use of analysis frameworks}

\myparagraph{Mitre \attack}
Subjects indicated Mitre \attack was the most used framework on their TH teams (see~\cref{framework_table}).
Some subjects said it was used to categorize tasks assigned to analysts.
Others said that only the intelligence component of the process used \attack to categorize indicators and detail adversary behaviors. 
One subject said that \attack was used only as an aid in describing behaviors written up in the final report. 
Since \attack is popular among government TH teams, TH teams that implement this framework will be more interoperable with other teams so that they can better tackle national security crises. 

\myparagraph{Pyramid of Pain}
Only one subject indicated they used the Pyramid of Pain but qualified their statement, saying that it was not in the current practice but in the \inlinequote{standard that we are creating} (this subject was a member of leadership).
Many experts recommend the use of Pyramid of Pain in cybersecurity\cite{davidjbianco_enterprise_2013,ec-council_what_2022}.
Building the Pyramid of Pain into a process will also assist in ensuring that the team does the maximum amount of damage to any adversaries that are discovered on a mission. 
Pyramid of Pain seems to lend itself especially well to government TH as it models how to do the most damage to an adversary's offensive capability.
Defending the US against such adversaries is one of CISA's explicit goals~\cite{noauthor_cisa_2022} and, to some extent, the goal of every government TH team.
We recommend using the Pyramid of Pain and indicate in \cref{fig:DetailedInducedDiagram} where it might fit into the process. 

\myparagraph{Kill Chain}
Kill chain was only used as far as it maps closely to the same categories in the Mitre \attack framework: \inlinequote{Our processes are loosely based on both [\attack and the Kill Chain]. 
The actual detailed hunting process is very closely related to the \attack framework.} 

\begin{table}
    \caption{Frameworks incorporated by subjects' process.}
    \begin{small}
    \renewcommand{\arraystretch}{0.85}
    \begin{tabularx}{\columnwidth}{c X}
        \toprule
        \textbf{Analysis Framework}  & \textbf{\# subjects who claim their process incorporated the framework (\# orgs.)} \\
        \toprule
        Mitre \attack  & 10 (2) \\
        Hypothesis Checking          & 7 (2) \\
        Lockheed Martin's Kill Chain & 3 (2) \\
        Pyramid of Pain              & 1 (1) \\
        \bottomrule
    \end{tabularx}
    \end{small}
    \label{framework_table}
\end{table}

\fi

\end{document}